\documentclass[trackchanges]{aastex701}

\usepackage{amsfonts,amssymb,amsthm,bbm,amsmath}
\usepackage{enumerate}
\usepackage{upgreek}
\usepackage{stackengine}
\usepackage{scalerel}
\usepackage{color,psfrag}
\usepackage{graphicx}
\usepackage{caption}
\usepackage{subcaption}
\usepackage{placeins}
\usepackage{booktabs}

\newcommand{\dd}{\mathrm{d}}
\begin{document}

\title{Kerr-Degenerate Shadows and Distinct Strong-Deflection Lensing in Rotating Hayward-like and Bardeen-like Geometries}

\author[gname=Chen-Hung, sname='Hsiao']{Chen-Hung Hsiao}
\affiliation{State Key Laboratory of Surface Physics, Center for Astronomy
and Astrophysics, Department of Physics, Center for Field Theory and Particle
Physics, and Institute for Nanoelectronic Devices and Quantum Computing,
Fudan University, Shanghai 200433, China}
\email[show]{chsiao@fudan.edu.cn}

\author[]{Limei Yuan}
\affiliation{State Key Laboratory of Surface Physics, Center for Astronomy
and Astrophysics, Department of Physics, Center for Field Theory and Particle
Physics, and Institute for Nanoelectronic Devices and Quantum Computing,
Fudan University, Shanghai 200433, China}
\email[show]{lmyuan24@m.fudan.edu.cn}

\author[]{Yidun Wan}
\affiliation{State Key Laboratory of Surface Physics, Center for Astronomy
and Astrophysics, Department of Physics, Center for Field Theory and Particle
Physics, and Institute for Nanoelectronic Devices and Quantum Computing,
Fudan University, Shanghai 200433, China}
\affiliation{Hefei National Laboratory, Hefei 230088, China}
\email[show]{ydwan@fudan.edu.cn}

\begin{abstract}
We study and compare rotating Hayward-like and Bardeen-like geometries within a common
Kerr-like axisymmetric ansatz, with the models distinguished by their
areal-radius functions $R_i(\rho)$.  On the adopted radial branches, the
positive minimum areal radius excludes the Kerr ring locus, and the curvature
invariants examined here remain finite.  We identify the parameter domains
containing two, one, or no horizons.  The two geometries have the same
asymptotic Komar charges but different finite-radius values.  If we treat the Einstein tensor as an effective stress tensor, then for every nonzero regularity length $\ell$, this ansatz violates the null and weak energy conditions, as demonstrated by the equatorial radial-null contraction.  For the complete photon family on a
black-hole branch, the shadow boundary is identical to that of Kerr at the
same mass, spin, and observer inclination, and is also identical
for the two regular geometries.  The corresponding area-equivalent shadow
diameters for M87* and Sgr~A* are consistent with the approximate Event Horizon
Telescope (EHT)-based
intervals adopted here, and we do not obtain further constraints on $\ell$.  By contrast, the prograde equatorial strong-deflection limit (SDL) calculations are only partly degenerate.  At fixed spin,
$u_m^+$ and $\theta_\infty^+$ are common to Kerr and the two regular geometries, whereas
$\bar a_+$, $\bar b_+$, the image separation, the relative magnitude, and the
subleading time-delay correction depend on $\ell$ and on the areal-radius
profile.  Within the stated strong-deflection approximation, future
measurements of $\Delta T_{2,1}$, $s^+$, and $r_{\rm mag}^+$ may help test
these rotating geometries and distinguish the Hayward-like and Bardeen-like
predictions from Kerr, including in the one-horizon parameter domain where the adopted branch contains no Kerr-like inner horizon.
\end{abstract}

\section{Introduction}
\label{sec:combined_introduction}

Gravitational lensing provides a way to test compact-object geometry in both
weak and strong gravitational fields.  In this work, we compare rotating
extensions of two static regular geometries constructed from Hayward-like and
Bardeen-like areal-radius profiles.  Our main question is whether their
shadows and strong-lensing observables can distinguish the two models from
Kerr and from each other.

\textbf{Principal results.}
We use a model label $i\in\{H,B\}$ for the Hayward-like and Bardeen-like
geometries.  Within the stated metric ansatz and adopted radial branches, our
main results are as follows.
\begin{enumerate}
\item \textit {Neither rotating black hole exhibits the ring singularity found in the Kerr black hole.} We identify the parameter domains containing two, one, or
no horizons without assuming a global extension through the branch endpoint.
The two models have the same asymptotic Komar charges but different
finite-radius profiles.  If the Einstein tensor is interpreted as an
effective source, an equatorial radial-null contraction, shows local NEC
violation, and therefore WEC violation, for every nonzero regularity length $\ell$ in
the strict exterior.
\item \textit{The celestial
shadow boundary for both black holes is identical to that of Kerr at the same mass, spin, and
observer inclination} 
 For the complete photon family on a black-hole branch, the calculated ideal shadow
sizes of rotating Hayward-like and rotating Bardeen-like for M87* and Sgr~A* lie within the approximate EHT-based intervals
adopted here, but this shadow-scale comparison does not constrain $\ell$. 
\item \textit{The prograde equatorial strong-deflection predictions are only partly
degenerate.}  At fixed spin, $u_m^+$, $\theta_\infty^+$, and the leading
time-delay contribution $2\pi u_m^+$ are common to Kerr and the two regular
models.  In contrast, $\bar a_+$, $\bar b_+$, the image separation $s^+$,
the relative magnitude $r_{\rm mag}^+$, and the subleading time-delay
correction depend on $\ell$ and differ between the Hayward-like and
Bardeen-like geometries.  These differences can distinguish the models in
principle within the stated SDL approximation.
\end{enumerate}

\textbf{Motivation and background.}
Gravitational lensing is the deflection of light by a gravitating object and
is now a standard tool for studying mass distributions and spacetime geometry
\cite{Wambsganss:1998gg,Bartelmann:2016dvf}.  In the weak-field regime, the
deflection angle can be obtained from perturbative null geodesics or from the
optical-geometry formulation based on the Gauss--Bonnet theorem
\cite{Gibbons:2008rj,Werner:2012rc}.  The weak-field approximation ceases to be
adequate when a photon passes close to an unstable photon orbit, where the
bending angle becomes large and multiple-loop trajectories can occur.

The foundations of black-hole strong lensing were developed over several
stages.  Darwin analyzed Schwarzschild photon trajectories near the critical
orbit and identified the sequence of highly deflected images
\cite{10.1098/rspa.1959.0015}.  Virbhadra and Ellis later formulated
Schwarzschild black-hole lensing in terms of relativistic images
\cite{Virbhadra:1999nm}, while Luminet calculated the visual appearance of a
Schwarzschild black hole surrounded by a thin accretion disk
\cite{1979A&A....75..228L}.  Exact integral lens equations were developed by
Frittelli, Kling, and Newman~\cite{PhysRevD.61.064021}.  Bozza subsequently
organized the near-critical logarithmic divergence into the
strong-deflection-limit (SDL) formalism and related it to observable image
positions and magnifications~\cite{Bozza:2001xd,Bozza:2002zj}.  The same
framework also provides standard approximations for lens equations and
relativistic-image time delays~\cite{Bozza:2003cp,Bozza:2008ev}.

In geometric optics, the unstable photon-orbit family also determines the
boundary of the ideal black-hole shadow
\cite{Bardeen:1973shadow,Perlick:2021aok}.  The shadow and the SDL images probe
related but not identical information: the celestial critical curve is fixed
by critical impact parameters, whereas image separations, magnifications, and
travel times also depend on photon propagation away from the critical orbit.
The horizon-scale images of M87* and Sgr~A* obtained by the EHT has made comparisons between strong-field calculations and
observations especially relevant
\cite{EventHorizonTelescope:2019dse,EventHorizonTelescope:2019ggy,
EventHorizonTelescope:2022apq,EventHorizonTelescope:2022wkp,
EventHorizonTelescope:2022xqj}.  These observations reconstruct plasma
emission around the compact objects, so the bright emission ring is not
identical to the geometric critical curve of a vacuum null-geodesic
calculation.  We therefore use the EHT results as a shadow-scale consistency
check rather than as a direct measurement of individual SDL images.

Regular-black-hole geometries provide phenomenological settings in which the
central curvature singularity of the classical solution is replaced by a
regular core.  The Bardeen and Hayward metrics are two standard static
examples~\cite{1968qtr..conf...87B,Hayward:2005gi,AyonBeato:1998ub}.
Strong-deflection lensing has been investigated for these and other static
regular black holes, as well as for several rotating nonsingular geometries
\cite{Eiroa:2010wm,Zhao:2017cwk,regularbh1,regularbh2,rotatingbh1,
Xie:2024srr,Guo:2025rrbl}.

A recurring issue is that many regular models contain an inner Cauchy horizon,
which marks the boundary of the region uniquely determined by initial data.
Extending the spacetime beyond this surface can therefore lead to a loss of
predictivity.  Perturbative studies further indicate that Cauchy horizons may
be subject to mass inflation or related instabilities
\cite{Penrose:1969pc,Poisson:1990eh,Maeda:2005yd,Brown:2011tv,
Carballo-Rubio:2018pmi}.  The construction of Ref.~\cite{Calza:2025mrt}
instead introduces Hayward-like and Bardeen-like areal-radius functions whose
adopted static branches begin at a nonzero minimum radius and exclude the
usual static inner root.  Rotation modifies this picture because the
Kerr-like inner root is positive. Depending on spin and regularity length $\ell$,
the adopted branch can contain two horizons, only an outer horizon, or no
horizon.  The static no-Cauchy-horizon argument therefore applies directly
only in the one-horizon rotating domain.

Extending a generic static regular geometry to rotation is not unique.  The
resulting spacetime must be checked independently for curvature regularity,
horizon structure, energy conditions, and its physical domain
\cite{Bambi:2013ufa,Neves:2014aba,Azreg-Ainou:2014pra,
Torres:2022regularreview}.  We consequently treat the rotating metrics below
as effective geometries and do not assign them a unique microscopic matter
source.  The static lensing properties of the Hayward-like and Bardeen-like
black holes used here were studied separately in
Refs.~\cite{Hsiao:2026oti,Yuan:2026bardeen}.  The present comparison places
their rotating extensions in the same conventions and on the same parameter
grid.  Because the two geometries have the same Kerr-like dependence on the
areal radius but different relations between areal and coordinate radius,
they provide a direct test of which observables depend only on the shadow
boundary and which retain information about the regularity length $\ell$.

Section~\ref{sec:combined_axisymmetric_extension} introduces the rotating
geometries and analyzes their regularity, horizons, Komar charges, effective
sources, and shadows.  Sections~\ref{sec:combined_lensing_basics} and
\ref{sec:combined_rotating_sdl} develop the equatorial lensing and SDL
formulas.  The relativistic-image, time-delay, and EHT comparisons are given
in Section~\ref{sec:combined_rotating_observables}, and the main
assumptions, limitations, and conclusions are summarized in
Section~\ref{sec:combined_conclusions}.

\subsection{The two nonrotating seeds}
\label{subsec:combined_static_seeds}

Both static geometries can be written in the common form
\cite{Calza:2025mrt}
\begin{equation}
 ds_i^2=-F_i(\rho)\mathrm{d}t^2+\frac{\mathrm{d}\rho^2}{F_i(\rho)}
 +R_i^2(\rho)\mathrm{d}\Omega^2,
 \qquad
 F_i(\rho)=1-\frac{2m}{R_i(\rho)},
\label{eq:combined_static_metric}
\end{equation}
where $m$ is the geometric mass and $\ell$ is a nonnegative regularity length
treated phenomenologically.
The physical exterior branch starts at the minimum of $R_i(\rho)$ and is
restricted to $R_i'(\rho)\geq0$.  The two areal-radius functions are
\begin{align}
 R_H(\rho)&=\rho+\frac{2m\ell^2}{\rho^2},
 \label{eq:combined_hayward_map}\\
 R_B(\rho)&=\frac{(\rho^2+\ell^2)^{3/2}}{\rho^2}.
 \label{eq:combined_bardeen_map}
\end{align}
For the Hayward-like seed, Eq.~\eqref{eq:combined_static_metric} is
equivalent to $F_H=1-2M_H(\rho)/\rho$ with
\begin{equation}
 M_H(\rho)=\frac{m\rho^3}{\rho^3+2m\ell^2}
\end{equation}
because $M_H/\rho=m/R_H$.  The Bardeen-like function can similarly be written
with $M_B(\rho)=m\rho^3/(\rho^2+\ell^2)^{3/2}$.  Thus, both lapse functions
take the Schwarzschild-like form as functions of $R_i$, although the radial
line element is not a mere coordinate rewriting of Schwarzschild when
$R_i'\neq1$. The branch data are summarized in Table~\ref{tab:combined_static_maps}.
\begin{table*}[t!]
\centering
\caption{Areal-radius functions and branch minima of the two nonrotating seeds.}
\label{tab:combined_static_maps}
\renewcommand{\arraystretch}{1.35}
\begin{tabular}{ccccc}
\toprule
Model & $R_i(\rho)$ & $\rho_{0,i}$ & $R_{0,i}=R_i(\rho_{0,i})$
& $R_i'(\rho)$ \\
\midrule
Hayward-like
& $\displaystyle \rho+\frac{2m\ell^2}{\rho^2}$
& $\displaystyle (4m\ell^2)^{1/3}$
& $\displaystyle \frac{3}{2}(4m\ell^2)^{1/3}$
& $\displaystyle 1-\frac{4m\ell^2}{\rho^3}$ \\
Bardeen-like
& $\displaystyle \frac{(\rho^2+\ell^2)^{3/2}}{\rho^2}$
& $\displaystyle \sqrt{2}\,\ell$
& $\displaystyle \frac{3\sqrt{3}}{2}\ell$
& $\displaystyle
\frac{\sqrt{\rho^2+\ell^2}(\rho^2-2\ell^2)}{\rho^3}$ \\
\bottomrule
\end{tabular}
\end{table*}

The static horizon condition is $R_i(\rho_h)=2m$.  A horizon belongs to the
adopted branch only if $R_{0,i}<2m$.  Both functions give the same
dimensionless bound,
\begin{equation}
 \frac{\ell}{m}<\frac{4}{3\sqrt{3}}.
\label{eq:combined_static_bound}
\end{equation}
At equality, the formal Schwarzschild-like root reaches the boundary of the
adopted branch.  We do not classify that one-sided boundary point as a horizon
without constructing an extension through it; larger values are horizonless.
The two functions nevertheless differ asymptotically:
\begin{align}
 R_H(\rho)&=\rho+\mathcal O(\rho^{-2}),\notag\\
 R_B(\rho)&=\rho+\frac{3\ell^2}{2\rho}
 +\mathcal O(\rho^{-3}).
\end{align}
This difference is invisible to the algebraic horizon equation in $R_i$ but
is retained by radial function.

\section{Axisymmetric Extension and Geometric Regularity}
\label{sec:combined_axisymmetric_extension}

We adopt the real rotating prescription of
Azreg-A\"{\i}nou \cite{Azreg-Ainou:2014aqa,Azreg-Ainou:2014pra}.  Unlike the
original Newman--Janis algorithm, it does not require an ambiguous complex
extension of the functions defining the nonrotating metric
\cite{Newman:1965tx,Drake:1998gf}.  Instead, it directly provides a
Boyer--Lindquist-like rotating ansatz that recovers the static seed when
$a=0$.  Its Kerr-like radial-profile form is related to the
G\"{u}rses--G\"{u}rsey class and has been widely used for rotating regular
geometries \cite{Gurses:1975vu,Bambi:2013ufa,Toshmatov:2014nya,
Neves:2014aba,Torres:2016gpe}.  We treat the resulting metrics as effective
nonvacuum geometries; the prescription does not select a unique underlying
matter model.  For either static seed, define
\begin{equation}
 \Sigma_i(\rho,\theta)=R_i^2(\rho)+a^2\cos^2\theta,
 \qquad
 \Delta_i(\rho)=R_i^2(\rho)-2mR_i(\rho)+a^2,
\label{eq:combined_sigma_delta}
\end{equation}
and $2f_i(\rho)=2mR_i(\rho)$.  The resulting Boyer--Lindquist-like line
element is
\begin{align}
 \dd s_i^2={}&-\left(1-\frac{2mR_i}{\Sigma_i}\right)\dd t^2
+\frac{\Sigma_i}{\Delta_i}\dd \rho^2+\Sigma_i \dd \theta^2
-\frac{4amR_i\sin^2\theta}{\Sigma_i}\dd t\,\dd \phi\notag\\
&+\frac{(R_i^2+a^2)^2-a^2\Delta_i\sin^2\theta}{\Sigma_i}
 \sin^2\theta\,\dd\phi^2.
\label{eq:combined_rotating_metric}
\end{align}
Here and below, an unlabeled $R_i$ in a metric expression means
$R_i(\rho)$.  Setting $a=0$ returns the corresponding static seed, while
$\ell=0$ gives $R_i=\rho$ and recovers Kerr.

A central feature of this extension is the avoidance of the Kerr ring locus
on the adopted radial branch.  The radial domain is restricted
to $\rho\geq\rho_{0,i}$, where the areal-radius function has the strictly positive
minimum $R_i(\rho)\geq R_{0,i}>0$.  Therefore,
\begin{equation}
 \Sigma_i(\rho,\theta)=R_i^2(\rho)+a^2\cos^2\theta
 \geq R_{0,i}^2>0,
\end{equation}
and the Kerr ring locus $\Sigma=0$ is not part of that domain.  The curvature
invariants, including the Kretschmann scalar
$\mathcal K=R_{\mu\nu\alpha\beta}R^{\mu\nu\alpha\beta}$, remain finite throughout
the adopted branch, showing that it contains no scalar-curvature singularity.
This does not by itself establish that geodesics can be continued through the
minimum-radius endpoint.  A root of $\Delta_i$ at
$\rho_h>\rho_{0,i}$ defines a horizon on the open physical branch.  By
contrast, when the root occurs at $\rho_{0,i}$, it lies at the
boundary of the adopted spacetime and cannot be identified as a genuine
horizon without first constructing a smooth extension through that boundary.

For $\ell>0$, the rotating geometries are generally not Ricci-flat: the Ricci
scalar $\mathcal R$ and $R_{\mu\nu}R^{\mu\nu}$ need not vanish.  Through
$G_{\mu\nu}=8\pi T^{\rm eff}_{\mu\nu}$, they define an effective, generally
anisotropic source.  We use this effective description phenomenologically and
do not assume a particular microscopic matter Lagrangian.

This behavior differs from conventional rotating Hayward and Bardeen
extensions continued to $r=0$ \cite{Bambi:2013ufa,Azreg-Ainou:2014pra,
Torres:2016gpe}.  There, the Ricci scalar, Ricci-tensor square, and
Kretschmann scalar can remain bounded while approaching different finite
values depending on the way one approaches the origin.  In the present construction, that
locus is excluded by $R_{0,i}>0$.  For every $\ell>0$, the positive minimum
radius keeps $\Sigma_i$ nonzero on the adopted domain.  As $\ell\to0$, this
minimum approaches zero.  At $\ell=0$, the metric reduces to Kerr, and the
ring singularity at $\rho=0$, $\theta=\pi/2$ returns.

\subsection{Horizon structure and geometric domains}
\label{subsec:combined_horizons}

The horizons are determined by $\Delta_i(\rho)=0$.  Within the adopted domain,
the larger admissible root defines the outer horizon, while the smaller root,
when present, defines the inner horizon.  We do not automatically identify
the inner horizon as a Cauchy horizon because that classification requires a
global causal analysis.

For both areal-radius functions the horizon equation first takes the same Kerr-like
algebraic form,
\begin{equation}
 \Delta_i(\rho)=R_i^2(\rho)-2mR_i(\rho)+a^2=0.
\label{eq:combined_horizon_equation}
\end{equation}
It is therefore simplest to solve first for the radial profile $R_i$.  The two
formal roots are
\begin{equation}
 R_h^\pm=m\pm\sqrt{m^2-a^2},
 \qquad |a|\leq m,
\label{eq:combined_horizon_areal}
\end{equation}
and no real Kerr-like root exists when $|a|>m$.  On the adopted branch
$\rho\geq\rho_{0,i}$, each function $R_i(\rho)$ is one-to-one and increasing.
Consequently, every admissible $R_h^\pm$ determines a unique coordinate
location through
\begin{equation}
 R_i(\rho_{h,i}^\pm)=R_h^\pm.
\end{equation}
A root represents a horizon when $R_{0,i}<R_h^\pm$.  If
$R_{0,i}=R_h^\pm$, the endpoint is not treated as horizon here.  If $R_{0,i}>R_h^\pm$, the root lies below
the minimum of $R_i(\rho)$ and has no corresponding physical value of
$\rho$.

Let
\begin{equation}
 A=\frac{|a|}{m},\qquad \lambda=\frac{\ell}{m},\qquad
 x_h^\pm=1\pm\sqrt{1-A^2}.
\end{equation}
Equating the model-dependent branch minimum to $R_h^\pm$ gives
\begin{align}
 \lambda_{h,H}^\pm(A)&=\sqrt{\frac{2(x_h^\pm)^3}{27}},
 \label{eq:combined_hayward_horizon_thresholds}\\
 \lambda_{h,B}^\pm(A)&=\frac{2x_h^\pm}{3\sqrt{3}}.
 \label{eq:combined_bardeen_horizon_thresholds}
\end{align}
In particular, the two upper curves $\lambda_{h,i}^+(A)$ bound the parameter
region that retains an open-branch outer root.  Both start from
\begin{equation}
 \lambda_{h,H}^+(0)=\lambda_{h,B}^+(0)
 =\frac{4}{3\sqrt{3}}\simeq0.7698
\end{equation}
but decrease differently with spin.  In the limit $A\to1$,
\begin{equation}
 \lambda_{h,H}^+\longrightarrow\sqrt{\frac{2}{27}}\simeq0.2722,
 \qquad
 \lambda_{h,B}^+\longrightarrow\frac{2}{3\sqrt{3}}\simeq0.3849.
\label{eq:combined_extremal_spin_outer_bounds}
\end{equation}
For $0<A<1$, each model therefore has two open-branch horizons when
$0\leq \lambda<\lambda_{h,i}^-$, one outer horizon when
$\lambda_{h,i}^-<\lambda<\lambda_{h,i}^+$, and no horizon when
$\lambda>\lambda_{h,i}^+$.  Equality marks
the contact of a Kerr-like areal root with the branch minimum.  At $A=0$,
only the outer static horizon is present for
$\lambda<4/(3\sqrt{3})$.  At $A=1$, the two areal roots coincide at $R=m$.

The formal radii $R_h^\pm$ depend only on $m$ and $a$, whereas the minimum
allowed radius $R_{0,i}$ increases with $\ell$.  At
$R_{0,i}=R_h^\pm$, the corresponding root reaches the minimum-radius endpoint;
for larger $\ell$, it lies below the allowed range and no longer represents a
horizon.  This is different from extremality: the two roots merge only at
$A=1$.  The merged root is an extremal horizon when $R_{0,i}<m$, while
$R_{0,i}=m$ is endpoint contact and is not classified as a horizon here.
In the one-horizon domain, $R_h^-<R_{0,i}<R_h^+$, the inner root is absent
from the adopted domain and cannot play the usual Kerr-like Cauchy-horizon
role \cite{Calza:2025mrt}.  In the two-horizon domain, both roots are present.
Determining whether the smaller root is globally a Cauchy horizon, and
assessing its stability, require a separate causal-extension analysis.

\begin{figure*}[t!]
\centering
\includegraphics[width=0.96\textwidth]
{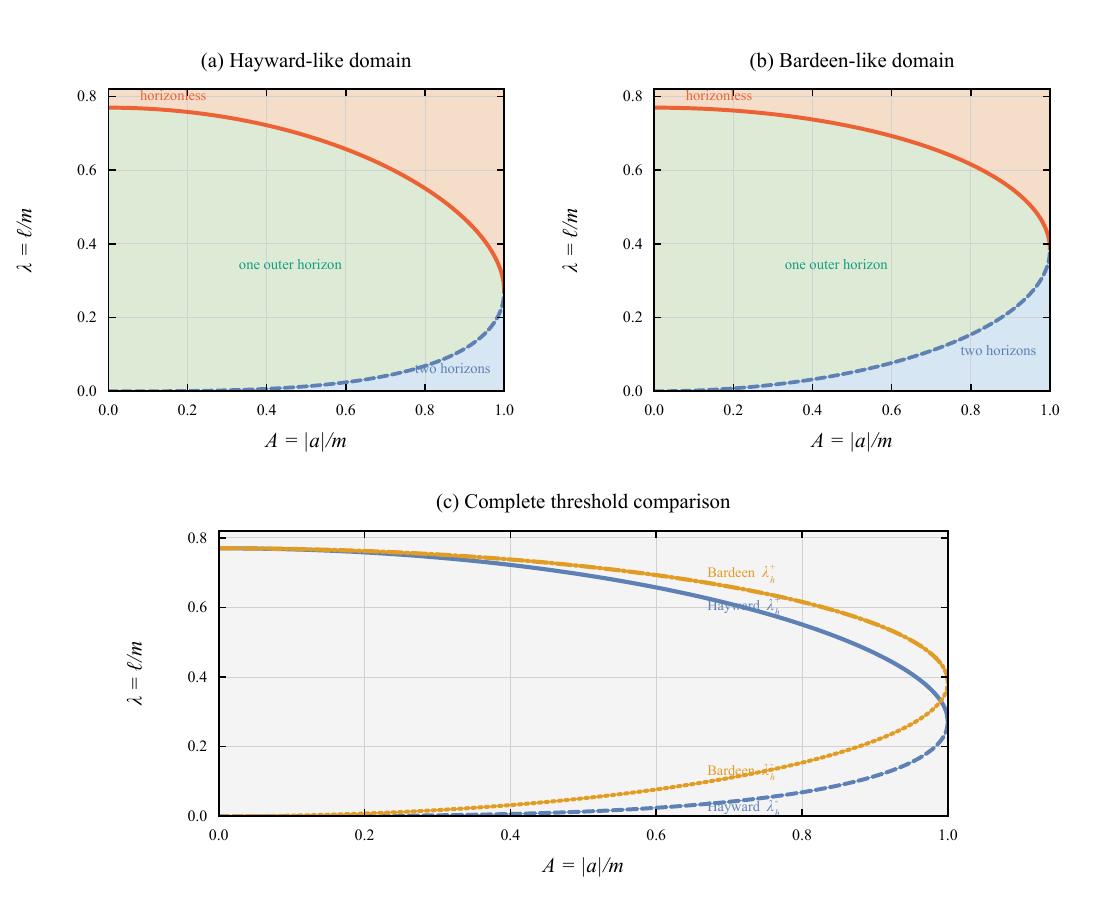}
\caption{Horizon domains in the dimensionless plane
$A=|a|/m$, $\lambda=\ell/m$.  Panels (a) and (b) show the analytic Hayward-like and
Bardeen-like thresholds: the dashed curve is $\lambda_{h,i}^{-}$ and the solid
curve is $\lambda_{h,i}^{+}$.  For $0<A<1$ and away from the threshold curves,
the regions below, between, and above them contain two, one, and no
adopted domain roots with the corresponding local horizon interpretation,
respectively.  Points on a threshold
represent formal contact with the adopted domain endpoint and are not assigned a
separate horizon class here.  The axes $A=0$ and $A=1$ obey the exceptional
cases described in the text.  Panel (c) overlays all four thresholds: color identifies the
geometry, while the dashed and solid curves identify the lower and upper
thresholds.  For $0<A<1$, $\lambda_{h,B}^{+}>\lambda_{h,H}^{+}$, so a finite parameter
range admits a Bardeen-like outer horizon but no Hayward-like outer horizon.
The two outer bounds meet at $A=0$, where both equal
$4/(3\sqrt{3})$.}
\label{fig:combined_horizon_parameter_domains}
\end{figure*}

\begin{figure*}[t!]
\centering
\includegraphics[width=0.94\textwidth]
{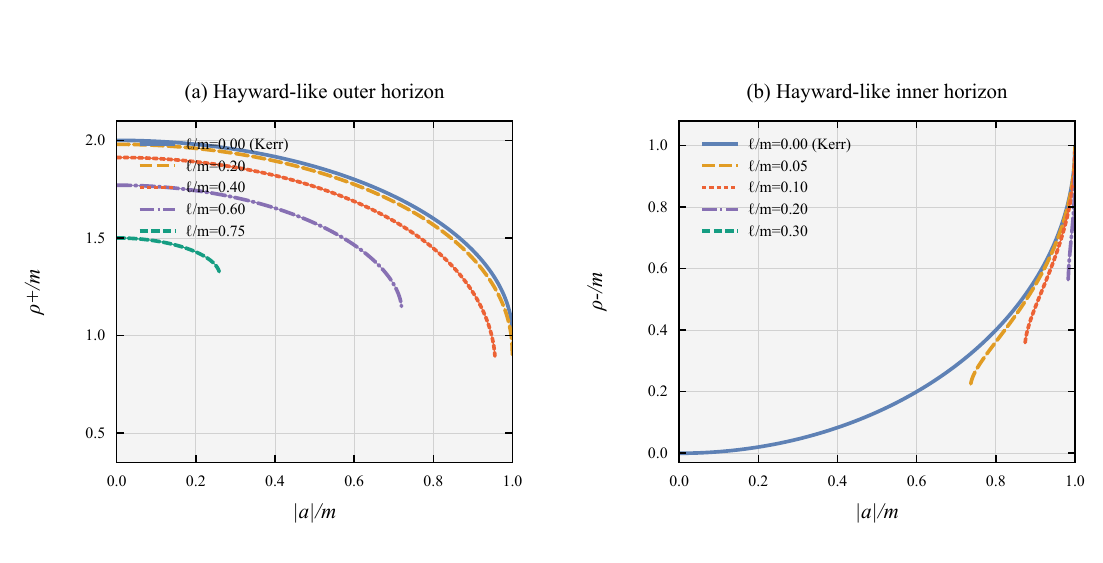}
\caption{Coordinate locations of the Hayward-like outer and inner
open-branch horizons compared with Kerr (blue curves).  A non-Kerr
curve terminates at the limiting contact of its corresponding areal root with
$R_{0,H}$; the endpoint itself is not classified as a horizon.  The inner root exists on the
adopted branch only in the small-$\ell$ two-root wedge.  Its global
identification as a Cauchy horizon is not assumed here.}
\label{fig:combined_hayward_horizon_coordinates}
\end{figure*}

\begin{figure*}[t!]
\centering
\includegraphics[width=0.94\textwidth]
{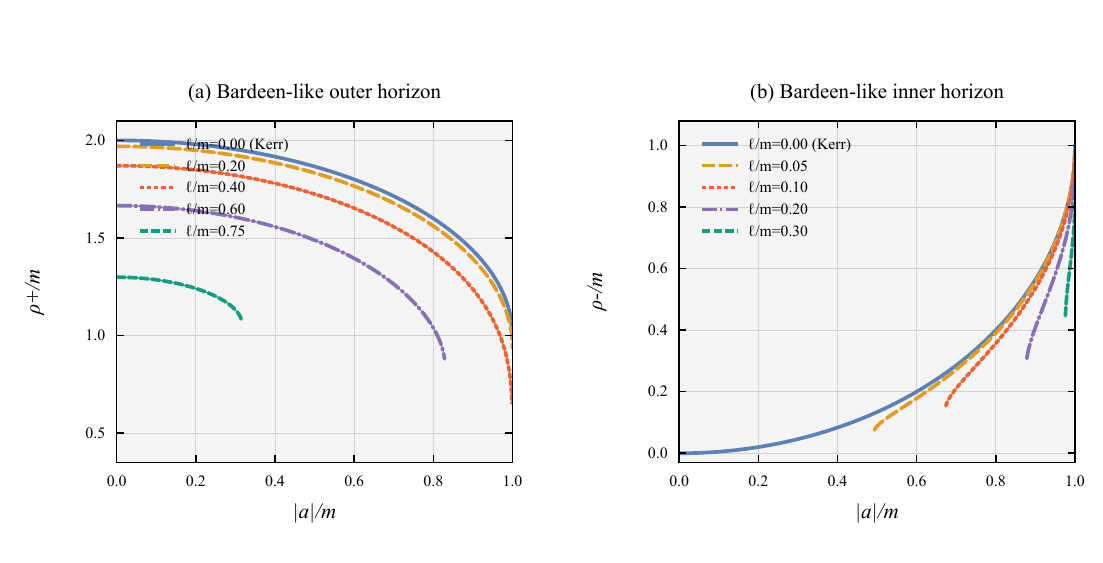}
\caption{Coordinate locations of the Bardeen-like open-branch horizons as
functions of spin.  The left and right panels show $\rho_{+,B}$
and $\rho_{-,B}$, respectively.  Each non-Kerr curve terminates at the
limiting contact of its Kerr-like areal root with $R_{0,B}$; the endpoint
itself is not classified as a horizon.  A missing continuation means that the
root lies below the minimum of the adopted domain.  The $\ell=0$
curves reproduce Kerr.}
\label{fig:combined_bardeen_horizon_structure}
\end{figure*}

Figures~\ref{fig:combined_horizon_parameter_domains}--
\ref{fig:combined_bardeen_horizon_structure} show the same horizon
classification in a common plotting convention.  These parameter-domain
figures display both lower and upper branch-contact thresholds and directly
compare the two outer-horizon bounds.  

\FloatBarrier

\subsection{Finite-surface Komar charges}
\label{subsec:combined_komar}

The Komar formalism provides a conserved definition of mass and angular momentum for stationary, axisymmetric spacetimes, such as rotating black holes. Let $S_\rho$ be the oriented two-surface at constant $t$ and $\rho$, with
stationary and axial Killing fields $\xi=\partial_t$ and
$\psi=\partial_\phi$.  For signature $(-,+,+,+)$ and orientation
$dt\wedge d\rho\wedge d\theta\wedge d\phi>0$, the Komar mass and angular
momentum are defined by \cite{Komar:1959,Wald:1984gr,Ali:2024rqcbh}
\begin{equation}
 M_K=-\frac{1}{8\pi}\int_{S_\rho}\nabla^\mu\xi^\nu \mathrm{d}S_{\mu\nu},
 \qquad
 J_K=\frac{1}{16\pi}\int_{S_\rho}\nabla^\mu\psi^\nu \mathrm{d} S_{\mu\nu}.
\label{eq:combined_komar_tensor_form}
\end{equation}
 Here, $dS_{\mu\nu}=2n_{[\mu}s_{\nu]}dA$ the oriented binormal area element of $S_\rho$, $n^\mu$ is the future-directed timelike unit normal, $s^\mu$ is the outward spacelike unit normal, and $dA$ is the proper area element. The relative factors $1/(8\pi)$ and $1/(16\pi)$ are the standard
Komar normalizations in units $G=c=1$.
Evaluation of the complete stationary $t$--$\phi$ block gives the following
common form within the adopted ansatz:
\begin{equation}
 M_{K,i}[S_\rho]=mR_i'(\rho),
 \qquad
 J_{K,i}[S_\rho]=amR_i'(\rho)=aM_{K,i}[S_\rho].
\label{eq:combined_komar_closed}
\end{equation}
Their explicit normalized profiles are
\begin{align}
 \frac{M_{K,H}}{m}=\frac{J_{K,H}}{am}
 &=1-\frac{4m\ell^2}{\rho^3},
 \label{eq:combined_hayward_komar}\\
 \frac{M_{K,B}}{m}=\frac{J_{K,B}}{am}
 &=\frac{\sqrt{\rho^2+\ell^2}(\rho^2-2\ell^2)}{\rho^3}.
 \label{eq:combined_bardeen_komar}
\end{align}
For $\ell>0$, the adopted radial domains are
\begin{equation}
 \rho\geq\rho_{0,H}=(4m\ell^2)^{1/3},
 \qquad
 \rho\geq\rho_{0,B}=\sqrt{2}\,\ell,
\end{equation}
respectively.  Therefore, neither charge is evaluated at the formal
$\rho\to0$ behavior of the areal-radius functions.  At each branch minimum,
$R_i'(\rho_{0,i})=0$, and Eq.~\eqref{eq:combined_komar_closed} gives
\begin{equation}
 M_{K,i}[S_{\rho_{0,i}}]=J_{K,i}[S_{\rho_{0,i}}]=0.
\end{equation}
When $\ell=0$, both functions reduce to $R_i(\rho)=\rho$, and the geometry is Kerr.
Then $R_i'=1$, so $M_K=m$ and $J_K=am$ on every enclosed vacuum
surface.  For all $a$ and $\ell$, asymptotic flatness gives
\begin{equation}
 \lim_{\rho\to\infty}M_{K,i}[S_\rho]=m,
 \qquad
 \lim_{\rho\to\infty}J_{K,i}[S_\rho]=am.
\end{equation}
Finally, the
finite-radius dependence for $\ell>0$ is consistent with Komar
conservation: these geometries are not Ricci-flat, so Stokes' theorem relates the difference between charges on the effective matter
contained between two surfaces \cite{Wald:1984gr}.  In the vacuum Kerr limit, by contrast, the charges are independent of the chosen enclosing
surface.

\subsection{Effective source and weak energy condition}
\label{subsec:combined_wec}

For $\ell>0$, neither rotating geometry is Ricci-flat.  We define its
effective source by
\begin{equation}
 T^{\rm eff}_{\mu\nu}=\frac{G_{\mu\nu}}{8\pi}.
\end{equation}
The weak energy condition (WEC) is the invariant requirement
$T^{\rm eff}_{\mu\nu}v^\mu v^\nu\geq0$ for every timelike vector $v^\mu$.
The null energy condition (NEC) requires
$T^{\rm eff}_{\mu\nu}k^\mu k^\nu\geq0$ for every null vector $k^\mu$.
Because a null vector is the limit of timelike vectors and the stress tensor
is continuous, the WEC implies the NEC.  Therefore, one null direction with
$T^{\rm eff}_{\mu\nu}k^\mu k^\nu<0$ is sufficient to show the violation of both
conditions.  This one-way implication is why we use the NEC for the analytic
test: It gives an invariant scalar without requiring diagonalization of the
rotating $t$--$\phi$ stress block.  Conversely, a nonnegative result for one
null direction would not establish the WEC.  For completeness,
Appendix~\ref{app:nec_implies_wec} illustrates this implication using a family
of radially moving timelike observers.
When the stress tensor is Hawking--Ellis type I, this is equivalent to
$\varepsilon\geq0$ and $\varepsilon+P_j\geq0$ in its physical real
eigenframe~\cite{Hawking:1973uf,Kontou:2020bta}.  Rotation generally leaves a nonzero
$t$--$\phi$ energy flux in the Carter frame, so coordinate components or
unboosted diagonal entries cannot be identified directly with these
principal quantities.

A direct null contraction therefore avoids this diagonalization ambiguity.
In a stationary region $\Delta_i>0$, introduce the radial part of the Carter
orthonormal coframe
\cite{Carter:1968separable,Chandrasekhar:1983blackholes},
\begin{equation}
 e^{(0)}=\sqrt{\frac{\Delta_i}{\Sigma_i}}
 (\dd t-a\sin^2\theta\,\dd \phi),
 \qquad
 e^{(1)}=\sqrt{\frac{\Sigma_i}{\Delta_i}}\dd\rho,
\end{equation}
and choose the radial null vector $k^{(A)}=(1,1,0,0)$.  On the equatorial
plane, direct projection gives $G^i_{(0)(1)}=0$ and the model-independent
identity
\begin{equation}
 8\pi T^{{\rm eff},i}_{(A)(B)}k^{(A)}k^{(B)}
 =G^i_{(0)(0)}+G^i_{(1)(1)}
 =-\frac{2\Delta_i R_i''(\rho)}{R_i^3(\rho)}.
\label{eq:combined_universal_nec}
\end{equation}
For the two geometries,
\begin{align}
 R_H''(\rho)&=\frac{12m\ell^2}{\rho^4},
 \notag\\[-2mm]
 8\pi T^{{\rm eff},H}_{(A)(B)}k^{(A)}k^{(B)}
 &=-\frac{24m\ell^2\Delta_H}{\rho^4R_H^3}<0,
 \label{eq:combined_hayward_nec}\\
 R_B''(\rho)&=
 \frac{3\ell^2(\rho^2+2\ell^2)}
 {\rho^4\sqrt{\rho^2+\ell^2}},
 \notag\\[-2mm]
 8\pi T^{{\rm eff},B}_{(A)(B)}k^{(A)}k^{(B)}
 &=-\frac{6\ell^2\rho^2(\rho^2+2\ell^2)\Delta_B}
 {(\rho^2+\ell^2)^5}<0.
 \label{eq:combined_bardeen_nec}
\end{align}
Both $R_i''$ are strictly positive for $\ell>0$.  The displayed null
contraction is hence negative at every strict exterior equatorial point
where $\Delta_i>0$.  This supplies a sufficient local demonstration of the NEC,
and hence the WEC, violation within the effective-source interpretation. The same diagnosis is negative for
$a=0$, so the nonrotating seeds in
Eq.~\eqref{eq:combined_static_metric} also fail this WEC test for nonzero
$\ell$.  When $\ell=0$, $R_i''=0$ and Kerr vacuum is recovered.

\begin{figure*}[t!]
\centering
\includegraphics[width=0.94\textwidth]
{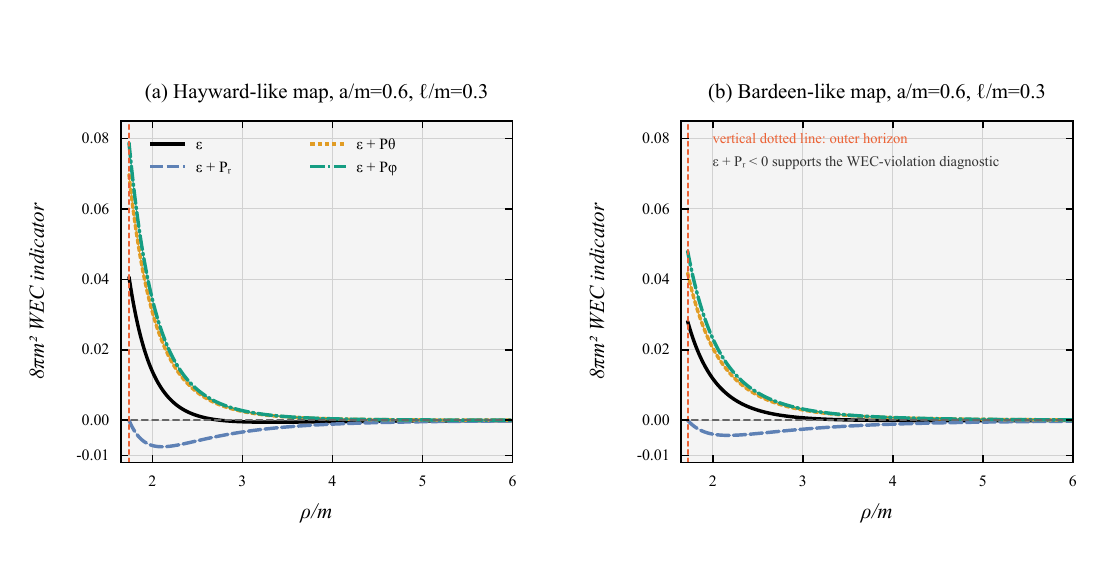}
\caption{Dimensionless type-I WEC indicators at $a/m=0.6$,
$\ell/m=0.3$, and $\theta=\pi/2$.  Both panels show $8\pi m^2$ times
$\varepsilon$ and $\varepsilon+P_j$, evaluated in the physical real
eigenframe of the mixed effective stress tensor.  The vertical dotted lines
mark the model-dependent outer horizons.  In each model,
$\varepsilon+P_r<0$ throughout the displayed strict exterior, in agreement
with Eqs.~\eqref{eq:combined_hayward_nec} and
\eqref{eq:combined_bardeen_nec}.}
\label{fig:combined_wec_profiles}
\end{figure*}

Figure~\ref{fig:combined_wec_profiles} also displays the physical
type-I principal-frame indicators for representative black holes.  The
negative $\varepsilon+P_r$ curve is consistent with the analytic
null-contraction test.  Its pointwise value need not coincide with
Eq.~\eqref{eq:combined_universal_nec} because the physical eigenframe
includes the boost required to remove the Carter-frame azimuthal energy
flux.  Similar violations occur in other rotating regular-black-hole
ansatzes~\cite{Bambi:2013ufa,Neves:2014aba}.

\FloatBarrier

Within the metric ansatz~\eqref{eq:combined_rotating_metric}, changing the
spin or reducing a nonzero $\ell$ does not change the sign of this particular
null contraction because the sign follows from $\Delta_i>0$ and
$R_i''>0$.  Whether a more general radial metric coefficient, rotating
completion, or matter model could avoid analogous violations, while
preserving regularity remains an open question.  The present calculation
does not establish that any of these modifications would succeed, and each
proposal would require a new invariant energy-condition analysis.

\subsection{Photon region and the black hole shadow}
\label{subsec:combined_shadow}
When light from distant sources passes near a black hole, some photon
trajectories escape to the observer whereas others are captured by the
horizon.  The boundary between these two outcomes forms the edge of the
black-hole shadow.  In a stationary axisymmetric spacetime, this boundary is
generated by unstable spherical photon orbits and therefore follows from the
null geodesic equations
\cite{Bardeen:1973shadow,Grenzebach:2014fha,Perlick:2021aok,
Ali:2024rqcbh}.  We accordingly determine the complete photon family before
projecting its critical trajectories onto the observer's screen.

Stationarity and axisymmetry give the conserved photon energy
$E=-p_t$ and axial angular momentum $L_z=p_\phi$.  As shown in
Appendix~\ref{app:combined_hj_separation}, the null Hamilton--Jacobi equation
also separates and introduces a Carter quantity $\mathcal Q$.  Define
\begin{equation}
 B_i(\rho)=R_i^2(\rho)+a^2.
\end{equation}
The corresponding impact parameters are
\begin{equation}
 \xi=\frac{L_z}{E},\qquad \eta=\frac{\mathcal Q}{E^2},
\end{equation}
and the separated potentials per unit energy are
\begin{align}
 \mathcal R_i(\rho)&=
 [B_i(\rho)-a\xi]^2
 -\Delta_i(\rho)\left[\eta+(a-\xi)^2\right],
\label{eq:combined_radial_potential}\\
 \frac{\Theta(\theta)}{E^2}
 &=\eta-\left(\xi^2\csc^2\theta-a^2\right)\cos^2\theta.
\label{eq:combined_angular_potential}
\end{align}
The corresponding first-order equations of motion are
\begin{equation}
 \left(\Sigma_i\frac{\dd\rho}{\dd\lambda}\right)^2
 =E^2\mathcal R_i(\rho),
 \qquad
 \left(\Sigma_i\frac{\dd\theta}{\dd\lambda}\right)^2
 =\Theta(\theta),
\label{eq:combined_photon_eom}
\end{equation}
where $\lambda$ is an affine parameter.  Thus $\mathcal R_i$ ($\Theta$) governs the
radial motion (polar motion).
An unstable spherical photon orbit at $\rho=\rho_p$ satisfies
\cite{Teo:2003bfn,Grenzebach:2014fha}
\begin{equation}
 \mathcal R_i(\rho_p)=0,\qquad
 \mathcal R_i'(\rho_p)=0,\qquad
 \mathcal R_i''(\rho_p)>0.
\label{eq:combined_spherical_conditions}
\end{equation}
The last sign follows from the convention in
Eq.~\eqref{eq:combined_radial_potential}; if
$V_{\rm eff}=-\mathcal R_i$, instability is equivalently
$V_{\rm eff}''<0$.

Solving the first two conditions yields the critical impact parameters for the unstable orbits
\begin{align}
 \xi_c(\rho_p)
 &=\frac{B_i\Delta_i'-2B_i'\Delta_i}
 {a\Delta_i'},
 \label{eq:combined_xi_general}\\
 \eta_c(\rho_p)
 &=\frac{4a^2(B_i')^2\Delta_i
 -\left(2B_i'\Delta_i-R_i^2\Delta_i'\right)^2}
 {a^2(\Delta_i')^2},
 \label{eq:combined_eta_general}
\end{align}
where all quantities are evaluated at $\rho_p$, and each prime denotes
$\dd/\dd\rho$.  The intermediate algebra is given in
Appendix~\ref{app:combined_hj_separation}.

With expressions
\begin{equation}
 B_i'=2R_iR_i',
 \qquad
 \Delta_i'=2(R_i-m)R_i',
\end{equation}
 where $R_i'>0$ in the adopted domain, the common factor $R_i'$ cancels out from
Eqs.~\eqref{eq:combined_xi_general} and
\eqref{eq:combined_eta_general}, leading to 
\begin{align}
 \xi_c(R_p)
 &=\frac{a^2(R_p+m)+R_p^2(R_p-3m)}
 {a(m-R_p)},
 \label{eq:combined_xi_kerr}\\
 \eta_c(R_p)
 &=\frac{R_p^3\left[4a^2m-R_p(R_p-3m)^2\right]}
 {a^2(m-R_p)^2},
 \label{eq:combined_eta_kerr}
\end{align}
where $R_p=R_i(\rho_p)$.
These reduce algebraically to the standard Kerr critical impact parameters
\cite{Bardeen:1973shadow,Chandrasekhar:1983blackholes,Teo:2003bfn}.

For $0<|a|<m$, the full shadow is generated by a continuous family of
spherical photon orbits~\cite{Teo:2003bfn,Perlick:2021aok},
\begin{equation}
 R_{\rm ph}^{\rm pro}\leq R_p\leq R_{\rm ph}^{\rm ret}.
\end{equation}
The equatorial prograde and retrograde endpoints have $\eta_c=0$ and are
\begin{align}
 \frac{R_{\rm ph}^{\rm pro}}{m}
 &=2\left[1+\cos\left(\frac{2}{3}
 \cos^{-1}\left(-\frac{|a|}{m}\right)\right)\right],
 \label{eq:combined_prograde_endpoint}\\
 \frac{R_{\rm ph}^{\rm ret}}{m}
 &=2\left[1+\cos\left(\frac{2}{3}
 \cos^{-1}\left(\frac{|a|}{m}\right)\right)\right].
 \label{eq:combined_retrograde_endpoint}
\end{align}
Intermediate values have $\eta_c>0$ and describe nonequatorial spherical
orbits whose polar coordinate oscillates.  The interval collapses to the
single Schwarzschild photon sphere $R_p=3m$ when $a\to0$.

To convert the photon-orbit family into an observable shadow curve, we place a
distant observer at an inclination $\theta_o$ measured from the rotation axis.
The celestial coordinates $X$ and $Y$ describe the apparent displacement on
the observer's screen perpendicular and parallel, respectively, to the
projected rotation axis.  They are obtained by projecting the photon momentum
onto a static orthonormal tetrad in the asymptotically flat region
\cite{Bardeen:1973shadow,Grenzebach:2014fha,Perlick:2021aok,
Ali:2024rqcbh}.  At leading order,
\begin{equation}
 p^{(t)}=E,\qquad
 p^{(\phi)}=\frac{L_z}{R_o\sin\theta_o},\qquad
 p^{(\theta)}=\frac{p_\theta}{R_o}.
\end{equation}
The Cartesian celestial coordinates are the physical transverse
displacements on a screen a distance $R_o$ away,
\begin{equation}
 X=-\lim_{R_o\to\infty}R_o\frac{p^{(\phi)}}{p^{(t)}},
 \qquad
 Y=\lim_{R_o\to\infty}R_o\frac{p^{(\theta)}}{p^{(t)}}.
\label{eq:combined_screen_definition}
\end{equation}
Using the impact parameters and
Eq.~\eqref{eq:combined_angular_potential} yields
\begin{equation}
 X=-\xi_c\csc\theta_o,
 \qquad
 Y=\pm\sqrt{\eta_c+a^2\cos^2\theta_o
 -\xi_c^2\cot^2\theta_o}.
\label{eq:combined_celestial_coordinates}
\end{equation}
The minus sign in $X$ is an orientation convention: positive axial angular
momentum appears on the opposite horizontal side of the observer's image
plane.  The two signs of $Y$ give the upper and lower halves of the shadow.

Equations~\eqref{eq:combined_xi_kerr}--
\eqref{eq:combined_celestial_coordinates} show that within geometric optics,
the shadow boundary of every black hole with the complete photon family in
the adopted domain coincides with that of Kerr at fixed $m$, $a$, and
$\theta_o$.
The regularity length changes only the coordinate radius obtained from
\begin{equation}
 R_H(\rho_{p,H})=R_p,
 \qquad
 R_B(\rho_{p,B})=R_p.
\label{eq:combined_coordinate_photon_radii}
\end{equation}
Thus, $\rho_{p,H}$ and $\rho_{p,B}$ generally differ even though their points
$(X,Y)$ coincide.

\begin{figure*}[t!]
\centering
\includegraphics[width=0.96\textwidth]
{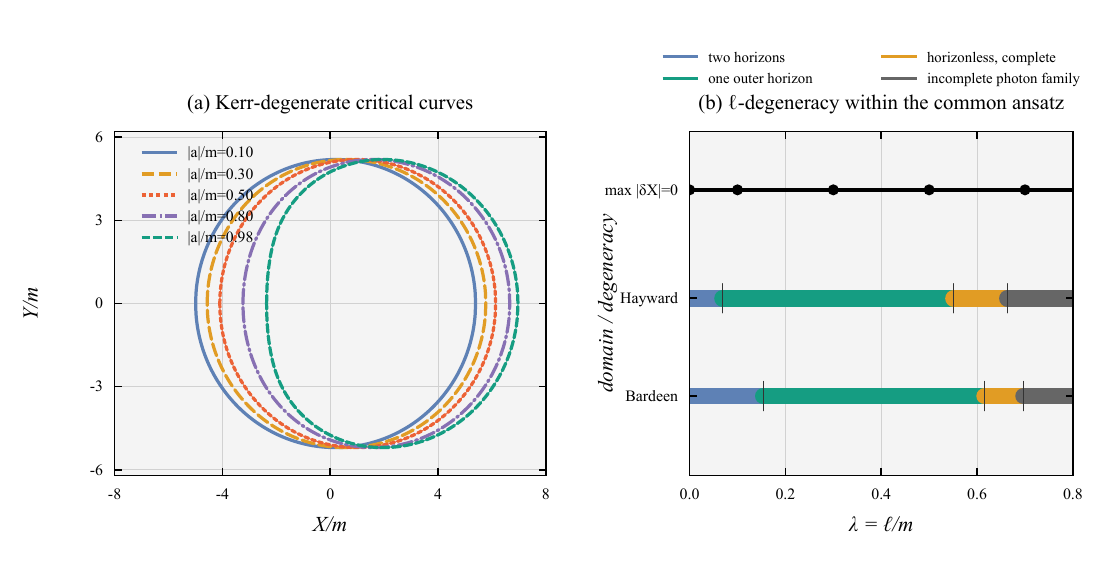}
\caption{Kerr-degenerate shadow boundaries within the common areal-radius
ansatz, for $m=1$ and an equatorial observer.  The left panel displays the complete
critical curve for
$|a|/m=0.1,0.3,0.5,0.8,0.98$.  For each spin, the Kerr, rotating
Hayward-like, and rotating Bardeen-like curves coincide.  At fixed
$|a|/m=0.8$, the right panel records the algebraically vanishing pointwise displacement
from Kerr and places the Hayward-like and Bardeen-like admissibility bars
under the same $\ell/m$ axis.  Blue, green, orange, and gray denote the
two-horizon, one-outer-horizon, horizonless-but-complete-photon-family, and
incomplete-photon-family regimes, respectively.  The transition marks are
formal endpoint-contact values and are not classified separately.  Thus
critical-curve degeneracy alone does not establish a black-hole
interpretation.}
\label{fig:combined_shadow_kerr}
\end{figure*}

For every black hole considered here, since
$R_{0,i}<R_h^+\leq R_{\rm ph}^{\rm pro}$, the entire spherical-photon
family lies on the physical domain outside the outer horizon.  For a formal
horizonless continuation, the formulas valid for $R_i'>0$ generate the complete
lensing critical curve only when
\begin{equation}
 R_{0,i}<R_{\rm ph}^{\rm pro}.
\end{equation}
At equality, the prograde endpoint coincides with the branch minimum, where
$R_i'=0$; the derivation of Eqs.~\eqref{eq:combined_xi_kerr} and
\eqref{eq:combined_eta_kerr}, which is divided by $R_i'$, does not apply.
Figure~\ref{fig:combined_shadow_kerr} illustrates this result for an
equatorial observer.  The left panel shows how spin shifts and flattens the
shadow boundary.  At each fixed spin, the Kerr, Hayward-like, and
Bardeen-like shadow boundaries coincide, so only one curve is visible.  The
right panel verifies numerically that varying $\ell$ produces no change in
the celestial coordinates within the plotted precision.  This curve
represents a black-hole shadow only when the chosen $(a,\ell)$ lies in the
one- or two-horizon domain defined by
Eqs.~\eqref{eq:combined_hayward_horizon_thresholds} and
\eqref{eq:combined_bardeen_horizon_thresholds}; for a horizonless parameter
choice, it is instead a lensing critical curve.

\FloatBarrier

The degeneracy is specific to observables determined only by the critical
impact parameters.  Radial propagation retains $\dd\rho$, $R_i'(\rho)$, and
the model-dependent relation between closest approach and areal radius.
Strong-deflection coefficients, relativistic-image separations,
magnifications, and subleading time delays remain sensitive to the choice of areal-radius function, even though their black-hole shadow boundaries coincide within this construction.

\section{Basics of Gravitational Lensing for a Rotating Lens}
\label{sec:combined_lensing_basics}

We now restrict the source, observer, and photon trajectory to the equatorial
plane.  This equatorial configuration allows the prograde and retrograde
photon trajectories to be treated separately in the strong-deflection
limit~\cite{Bozza:2001xd,Bozza:2002zj,Bozza:2008ev}.  The complete
off-equatorial shadow remains the two-dimensional celestial curve derived in
Section~\ref{subsec:combined_shadow}; the scalar formulas below instead
describe the prograde or retrograde sequence of equatorial relativistic
images.

For either model the equatorial metric has the common form
\begin{equation}
 \dd s_i^2=-\mathcal A_i(\rho)\dd t^2+\mathcal B_i(\rho)\dd\rho^2
 +\mathcal C_i(\rho)\dd\phi^2-\mathcal D_i(\rho)\dd t\,\dd \phi,
 \qquad g_{t\phi}=-\frac{\mathcal D_i}{2},
\label{eq:combined_lensing_metric}
\end{equation}
where
\begin{align}
 \mathcal A_i&=1-\frac{2m}{R_i},
 &\mathcal B_i&=\frac{R_i^2}{\Delta_i},\notag\\
 \mathcal C_i&=R_i^2+a^2+\frac{2ma^2}{R_i},
 &\mathcal D_i&=\frac{4am}{R_i}.
\label{eq:combined_lensing_functions}
\end{align}
The functions $\mathcal A_i$, $\mathcal C_i$, and $\mathcal D_i$ are Kerr-like
when expressed through $R_i$, but the radial measure
$\mathcal B_i(\rho)\dd\rho^2$ retains the model-dependent radial function.  This is why the
shadow boundary can be Kerr-degenerate while the bending integral is not.

Stationarity and axial symmetry give the conserved quantities
\begin{equation}
 E=\mathcal A_i\dot t+\frac{\mathcal D_i}{2}\dot\phi,
 \qquad
 L_z=-\frac{\mathcal D_i}{2}\dot t+\mathcal C_i\dot\phi,
 \qquad u\equiv\frac{L_z}{E}.
\label{eq:combined_lensing_EL}
\end{equation}
For a strictly equatorial orbit, $\eta=0$ and therefore $u=\xi$.  Solving
Eq.~\eqref{eq:combined_lensing_EL} and imposing the null condition gives
\begin{align}
 \dot t&=\frac{2E(2\mathcal C_i-\mathcal D_i u)}
 {4\mathcal A_i\mathcal C_i+\mathcal D_i^2},
 &
 \dot\phi&=\frac{2E(\mathcal D_i+2\mathcal A_i u)}
 {4\mathcal A_i\mathcal C_i+\mathcal D_i^2},\notag\\
 \dot\rho^{2}&=
 \frac{4E^2\mathcal P_i(\rho;u)}
 {\mathcal B_i(4\mathcal A_i\mathcal C_i+\mathcal D_i^2)},
 &
 \mathcal P_i(\rho;u)&\equiv
 \mathcal C_i-\mathcal D_i u-\mathcal A_i u^2.
\label{eq:combined_lensing_first_order}
\end{align}

At the closest approach $\rho_0$, $\dot\rho=0$, so
$\mathcal P_i(\rho_0;u)=0$.  The two signed roots are
\begin{equation}
 u_s(\rho_0)=
 \frac{-\mathcal D_{i0}+s
 \sqrt{\mathcal D_{i0}^2+4\mathcal A_{i0}\mathcal C_{i0}}}
 {2\mathcal A_{i0}},
 \qquad s=\pm1.
\label{eq:combined_impact_branches}
\end{equation}
For $a>0$ and $E>0$, the root with $u_+>0$ is prograde and the root with
$u_-<0$ is retrograde.  The rationalized prograde expression
\begin{equation}
 u_+(\rho_0)=
 \frac{2\mathcal C_{i0}}
 {\mathcal D_{i0}+\sqrt{\mathcal D_{i0}^2+
 4\mathcal A_{i0}\mathcal C_{i0}}}
\label{eq:combined_impact_prograde}
\end{equation}
remains finite at the equatorial stationary-limit surface
$\mathcal A_{i0}=0$.

Dividing $\dot\phi$ by $\dot\rho$ gives the equatorial bending integral
\begin{equation}
 \hat\alpha_s(\rho_0)=2\int_{\rho_0}^{\infty}
 \left|
 \frac{\sqrt{\mathcal B_i}(\mathcal D_i+2\mathcal A_i u_s)}
 {\sqrt{4\mathcal A_i\mathcal C_i+\mathcal D_i^2}
 \sqrt{\mathcal C_i-\mathcal D_i u_s-\mathcal A_i u_s^2}}
 \right|\dd\rho-\pi.
\label{eq:combined_exact_deflection}
\end{equation}
The absolute value defines a positive bending magnitude for either
orientation.  The factor of two assumes the usual asymptotic scattering
configuration, in which the ingoing and outgoing radial pieces contribute
equally.

For reference, the unapproximated Ohanian lens equation is~\cite{Bozza:2008ev}
\begin{equation}
 D_S\tan\beta=
 \frac{D_L\sin\theta-D_{LS}\sin(\hat\alpha_s-\theta)}
 {\cos(\hat\alpha_s-\theta)}.
\label{eq:combined_ohanian_lens_equation}
\end{equation}
Here, $\beta$ is the signed angular position of the unlensed source and
$\theta$ is the signed angular position of its lensed image, both measured
from the observer--lens optical axis in the equatorial plane.  The quantities
$D_L$, $D_{LS}$, and $D_S$ are the observer--lens, lens--source, and
observer--source angular-diameter distances.  A complete flux magnification
for a rotating lens is the inverse Jacobian determinant of the
two-dimensional lens map.  The scalar expressions used below refer only to
the equatorial image sequence; away from the equator, the transverse
derivative must be retained.

\section{Deflection Angle in the Rotating Strong Deflection Limit}
\label{sec:combined_rotating_sdl}

The exact bending integral in Eq.~\eqref{eq:combined_exact_deflection}
diverges when the turning point $\rho_0$ approaches an unstable equatorial
photon orbit.  Physically, a near-critical photon remains close to that orbit
for a long time and can wind around the black hole before escaping.
We use the standard construction and logarithmic expansion of
Refs.~\cite{Bozza:2001xd,Bozza:2002zj,Bozza:2003quasi}.

The shadow section uses $R_p$ to parametrize the complete family of spherical
photon orbits.  The prograde equatorial SDL selects one endpoint of that
family, where $\eta_c=0$:
\begin{equation}
 R_m^+
 \equiv R_p\big|_{\eta_c=0,\,\mathrm{pro}}
 =R_{\rm ph}^{\rm pro},
 \qquad
 \rho_{m,i}\equiv\rho_{p,i}^{\rm pro},
 \quad i\in\{H,B\}.
\label{eq:combined_shadow_sdl_notation}
\end{equation}
The associated impact parameter and coordinate radius obey
\begin{equation}
 u_m^+=u_{+,i}(\rho_{m,i}),\qquad
 R_i(\rho_{m,i})=R_m^+,\qquad i\in\{H,B\}.
\label{eq:combined_sdl_critical_data}
\end{equation}
Hence, $R_m^+$ and $u_m^+$ are the same for the two geometries and do not
depend on $\ell$.  Only their coordinate locations $\rho_{m,H}$ and
$\rho_{m,B}$ differ.

To isolate the divergent part, we follow the standard Bozza decomposition
and its rotating equatorial applications~\cite{Bozza:2002zj,
Bozza:2003quasi,rotatingbh1,rotatingbh2,regularbh1,loop1}.  The variable
\begin{equation}
 z=\frac{\mathcal A_i-\mathcal A_{i0}}{1-\mathcal A_{i0}},
\label{eq:combined_bozza_variable}
\end{equation}
maps the integration interval $\rho_0\leq\rho<\infty$ to $0\leq z<1$;
$z=0$ is the turning point, and $z\to1$ is spatial infinity.  The subscript
$0$ denotes evaluation at $\rho_0$. The deflection integral can be written as 
\begin{equation}
 I_i(\rho_0)=\hat\alpha_+(\rho_0)+\pi
 =\int_0^1 R_{{\rm Bz},i}(z,\rho_0)f_i(z,\rho_0)\dd z,   \end{equation}
where the integrand can be separated into a
regular factor $R_{{\rm Bz},i}$ and a factor $f_i$ containing the divergence,
\begin{align}
 R_{{\rm Bz},i}&=
 \frac{2(1-\mathcal A_{i0})\sqrt{\mathcal B_i}\,
 \mathcal A_{i0}(2\mathcal A_i u_++\mathcal D_i)}
 {\mathcal A_i'\sqrt{\mathcal C_i\mathcal A_{i0}}
 \sqrt{4\mathcal A_i\mathcal C_i+\mathcal D_i^2}},
\notag\\
 f_i&=\left[
 \mathcal A_{i0}-\mathcal A_i\frac{\mathcal C_{i0}}{\mathcal C_i}
 +\frac{u_+}{\mathcal C_i}
 (\mathcal A_i\mathcal D_{i0}-\mathcal A_{i0}\mathcal D_i)
 \right]^{-1/2}.
\label{eq:combined_bozza_functions}
\end{align}
where the unlabelled metric functions are evaluated at the radius
$\rho=\rho(z)$, and a prime denotes $d/d\rho$.  Near $z=0$,
\begin{equation}
 f_i(z,\rho_0)\simeq
 \frac{1}{\sqrt{\mathsf m_i(\rho_0)z+
 \mathsf n_i(\rho_0)z^2}},
\label{eq:combined_bozza_singular_factor}
\end{equation}
where $\mathsf m_i$ and $\mathsf n_i$ are the first two Taylor coefficients
of the expression under the square root.  At the critical orbit,
$\mathsf m_i(\rho_{m,i})=0$, and
$\mathsf n_{m,i}\equiv\mathsf n_i(\rho_{m,i})>0$
\cite{Bozza:2002zj,Bozza:2003quasi}.

The impact parameter is a function $u_+(\rho_0)$ of the turning point.
At the critical orbit, since $\dd u_+/\dd\rho_0=0$, the Taylor expansion of $u_+(\rho_0)$ begins at
quadratic order:
\begin{equation}
 u^{+}-u_m^+=\bar c_i(\rho_0-\rho_{m,i})^2+\cdots.
\label{eq:combined_critical_impact_expansion}
\end{equation}
Hence, $\bar c_i=\tfrac12u_+''(\rho_{m,i})$ is the leading coefficient relating
the impact parameter to the displacement of the turning point.
The resulting SDL expansion is
\begin{equation}
 \hat\alpha_+(u)=-\bar a_{+,i}
 \log\left(\frac{u}{u_m^+}-1\right)+\bar b_{+,i}
 +\mathcal O(u-u_m^+),
\label{eq:combined_sdl_expansion}
\end{equation}
with
\begin{align}
 \bar a_{+,i}&=\frac{R_{{\rm Bz},i}(0,\rho_{m,i})}
 {2\sqrt{\mathsf n_{m,i}}},
\notag\\
 \bar b_{+,i}&=-\pi+b_{D,i}+b_{R,i}
 +\bar a_{+,i}\log\left(
 \frac{\bar c_i\rho_{m,i}^2}{u_m^+}\right),
\label{eq:combined_bozza_coefficients}
\end{align}
where the divergent contribution $b_{D,i}$ is analytic, and $b_{R,i}$ is
the finite remainder after subtracting the critical singular term:
\begin{align}
 b_{D,i}&=2\bar a_{+,i}\log\left[
 \frac{2(1-\mathcal A_{im})}{\mathcal A_{im}'\rho_{m,i}}
 \right],\notag\\
 b_{R,i}&=\int_0^1\left[
 R_{{\rm Bz},i}(z,\rho_{m,i})f_i(z,\rho_{m,i})
 -R_{{\rm Bz},i}(0,\rho_{m,i})f_{0,i}(z,\rho_{m,i})
 \right]\dd z.
\label{eq:combined_bozza_finite_terms}
\end{align}
Here,
$f_{0,i}(z,\rho_{m,i})=1/(\sqrt{\mathsf n_{m,i}}\,z)$, the subscript
$m$ means evaluation at $\rho_{m,i}$, and $\log$ denotes the natural
logarithm.

For either geometry on the open physical branch, the leading coefficient is
\begin{equation}
 \bar a_{+,i}=\frac{\bar a_+^{\rm Kerr}}
 {R_i'(\rho_{m,i})}.
\label{eq:combined_abar_map_factor}
\end{equation}
This relation applies within the adopted domain where $R_i'>0$.
Since $R_i'\to1$ as $\ell\to0$, both geometries recover the Kerr coefficient.
The coefficient $\bar b_{+,i}$ also contains the finite integral $b_{R,i}$,
which is evaluated numerically.

\begin{figure*}[t!]
\centering
\includegraphics[width=0.98\textwidth]
{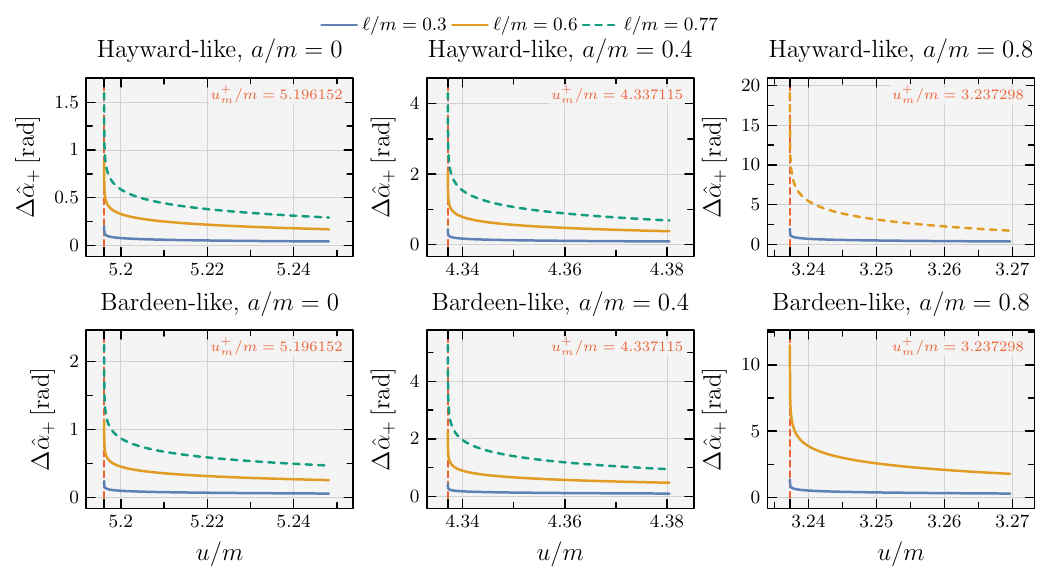}
\caption{SDL residual
$\Delta\hat\alpha_+=\hat\alpha_+(\ell)-\hat\alpha_+(0)$ on the same grid
as Fig.~\ref{fig:combined_sdl_deflection}.  The growing separation toward
$u_m^+$ comes from the model-dependent change in both $\bar a_+$ and
$\bar b_+$.}
\label{fig:combined_sdl_residuals}
\end{figure*}

Table~\ref{tab:combined_sdl_coefficients} compares both geometries on the same
parameter grid.  A dagger marks a parameter choice outside the black-hole
domain for which the exterior photon orbit is retained only as a formal
continuation.  A dash means that the minimum allowed radius removes the
usual exterior prograde critical orbit.
Figures~\ref{fig:combined_sdl_deflection} and
\ref{fig:combined_sdl_residuals} show the SDL calculation directly as a
function of $u/m$ and its deviation from Kerr.  They illustrate the
model dependence even though $u_m^+$ itself is unchanged.  Parameter choices outside
the black-hole domain are retained only as explicitly formal continuations.
\begin{table*}[t!]
\centering
\caption{Prograde equatorial SDL quantities.  The critical impact parameter
$u_m^+$ is common to both geometries.}
\label{tab:combined_sdl_coefficients}
\tabletypesize{\scriptsize}
\setlength{\tabcolsep}{3.2pt}
\renewcommand{\arraystretch}{1.00}
\begin{tabular}{ccccccccc}
\toprule
&&\multicolumn{3}{c}{Hayward-like}&\multicolumn{3}{c}{Bardeen-like}\\
\cmidrule(lr){4-6}\cmidrule(lr){7-9}
$a/m$ & $\ell/m$ & $u_m^+/m$
& $\rho_m/m$ & $\bar a_+$ & $\bar b_+$
& $\rho_m/m$ & $\bar a_+$ & $\bar b_+$\\
\midrule
0.0 & 0.00 & 5.196152 & 3.000000 & 1.000000 & -0.400230 & 3.000000 & 1.000000 & -0.400230 \\
 & 0.10 & 5.196152 & 2.997774 & 1.001487 & -0.402746 & 2.994990 & 1.001676 & -0.401834 \\
 & 0.30 & 5.196152 & 2.979727 & 1.013795 & -0.424130 & 2.954184 & 1.015835 & -0.416368 \\
 & 0.50 & 5.196152 & 2.942242 & 1.040866 & -0.474577 & 2.868271 & 1.048891 & -0.456894 \\
 & 0.77 & 5.196152 & 2.854467$^{\dagger}$ & 1.113547 & -0.632007 & 2.658556$^{\dagger}$ & 1.154160 & -0.640773 \\
\midrule
0.2 & 0.00 & 4.783246 & 2.759193 & 1.090302 & -0.426197 & 2.759193 & 1.090302 & -0.426197 \\
 & 0.10 & 4.783246 & 2.756561 & 1.092388 & -0.429769 & 2.753744 & 1.092465 & -0.428371 \\
 & 0.30 & 4.783246 & 2.735132 & 1.109829 & -0.460618 & 2.709211 & 1.110923 & -0.448399 \\
 & 0.50 & 4.783246 & 2.690100 & 1.149342 & -0.536872 & 2.614456 & 1.155411 & -0.507026 \\
 & 0.77 & 4.783246 & 2.581217$^{\dagger}$ & 1.264707 & -0.806444 & 2.375067$^{\dagger}$ & 1.313212 & -0.817276 \\
\midrule
0.4 & 0.00 & 4.337115 & 2.493363 & 1.220947 & -0.476140 & 2.493363 & 1.220947 & -0.476140 \\
 & 0.10 & 4.337115 & 2.490137 & 1.224118 & -0.481665 & 2.487330 & 1.223918 & -0.479319 \\
 & 0.30 & 4.337115 & 2.463708 & 1.251064 & -0.530630 & 2.437775 & 1.249656 & -0.509297 \\
 & 0.50 & 4.337115 & 2.407066 & 1.315254 & -0.661279 & 2.330624 & 1.314813 & -0.603543 \\
 & 0.77 & 4.337115 & 2.261508$^{\dagger}$ & 1.535867 & -1.245127 & 2.042916$^{\dagger}$ & 1.595935 & -1.259170 \\
\midrule
0.8 & 0.00 & 3.237298 & 1.811086 & 1.915896 & -0.952002 & 1.811086 & 1.915896 & -0.952002 \\
 & 0.10 & 3.237298 & 1.804947 & 1.929018 & -0.976863 & 1.802759 & 1.924801 & -0.963930 \\
 & 0.30 & 3.237298 & 1.752476 & 2.053232 & -1.234749 & 1.732587 & 2.008225 & -1.088847 \\
 & 0.50 & 3.237298 & 1.620740 & 2.504071 & -2.476433 & 1.565545 & 2.292820 & -1.673255 \\
 & 0.77 & 3.237298 & -- & -- & -- & -- & -- & -- \\
\bottomrule

\end{tabular}
\vspace{0.4em}
\begin{minipage}{\textwidth}
\centering
\includegraphics[width=0.88\textwidth]
{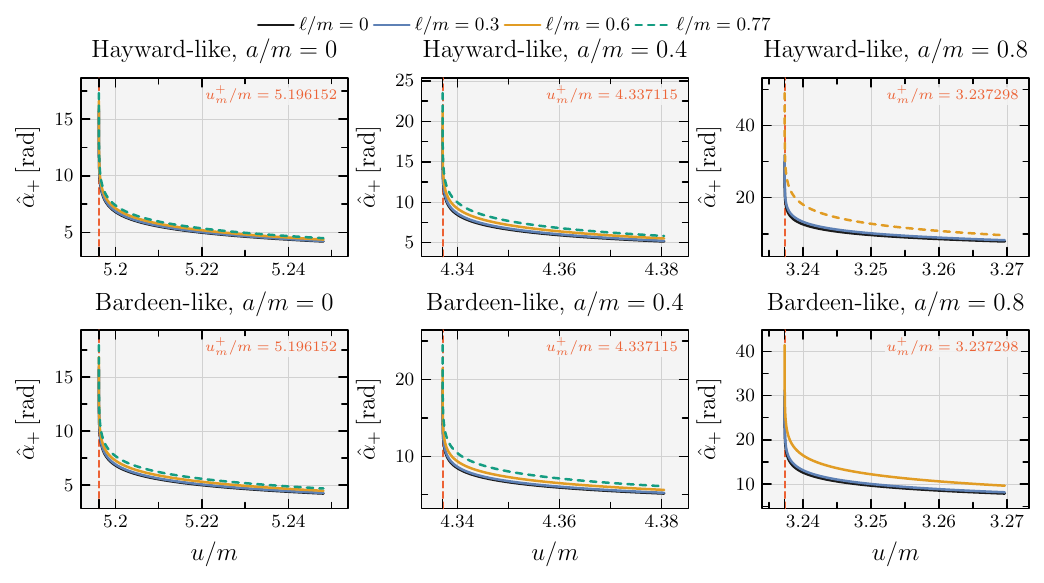}
\captionof{figure}{Prograde SDL bending angle from
Eq.~\eqref{eq:combined_sdl_expansion} for the Hayward-like (top) and
Bardeen-like (bottom) geometries at $a/m=0,0.4,0.8$.  Curves show
$\ell/m=0,0.3,0.6,0.77$; red vertical lines mark the listed $u_m^+$ values.
The dashed $\ell/m=0.77$ curves and the Hayward-like
$(a/m,\ell/m)=(0.8,0.6)$ curve are horizonless continuations.}
\label{fig:combined_sdl_deflection}
\end{minipage}
\end{table*}

\begin{figure*}[t!]
\centering
\includegraphics[width=\textwidth]
{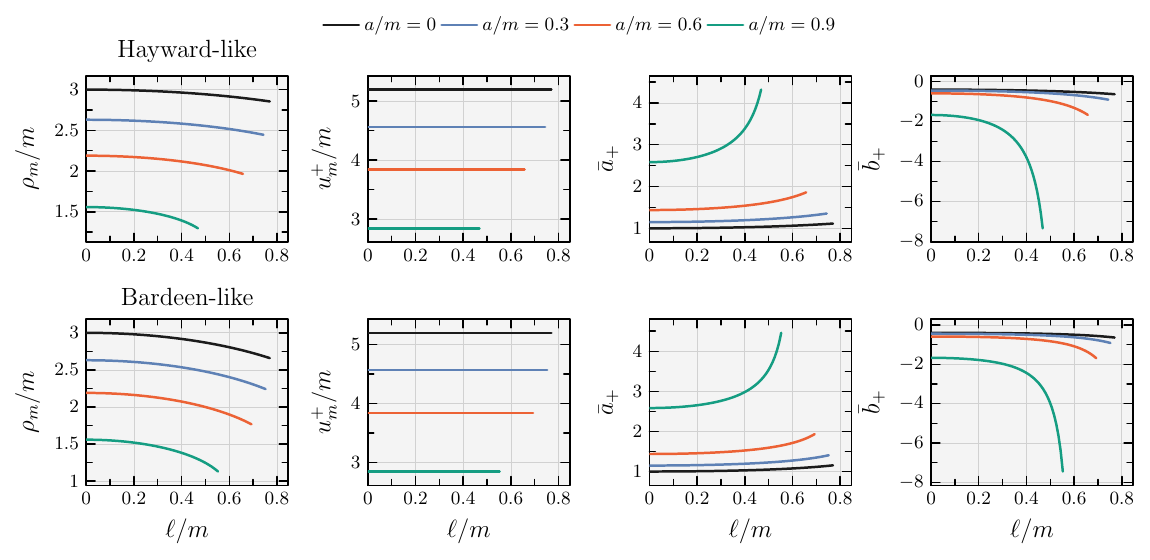}
\caption{Black-hole-only prograde SDL quantities for
$a/m=0,0.3,0.6,0.9$.  Every curve terminates at the corresponding
model-dependent outer-horizon bound.  At fixed spin, $u_m^+$ is constant,
whereas $\rho_m$, $\bar a_+$, and $\bar b_+$ differ between the two
geometries.}
\label{fig:combined_sdl_coefficients}
\end{figure*}
\FloatBarrier

\section{Rotating Strong-Field Images and Shadow Observables}
\label{sec:combined_rotating_observables}

\subsection{Relativistic images}

Photons with impact parameters just above $u_m^+$ can wind around the black
hole before reaching the observer, producing a sequence of prograde
relativistic images.  The positive integer $n=1,2,\ldots$ labels the winding
order, with $n=1$ and $n=2$ denoting the first and second relativistic images.  For a nearly aligned source, the deflection angle of the
$n$th image can therefore be written as
$\hat\alpha_+=2\pi n+\Delta\hat\alpha_n^+$, where
$|\Delta\hat\alpha_n^+|\ll1$.  Linearizing
Eq.~\eqref{eq:combined_ohanian_lens_equation} gives the equatorial thin-lens
equation~\cite{Bozza:2001xd,Bozza:2008ev}
\begin{equation}
 \beta=\theta-\frac{D_{LS}}{D_S}\Delta\hat\alpha_n^+.
\label{eq:combined_linear_lens_equation}
\end{equation}
The zeroth-order image position and its first correction are
\begin{align}
 \theta_{n,i}^{0,+}&=\frac{u_m^+}{D_L}\left(1+e_{n,i}^+\right),
 &
\theta_{n,i}^+&=\frac{u_m^+e_{n,i}^+D_S}
 {\bar a_{+,i}D_LD_{LS}}
 (\beta-\theta_{n,i}^{0,+}),
\label{eq:combined_relativistic_image_position}
\end{align}
where $e_{n,i}^+=\exp\left(
 \frac{\bar b_{+,i}-2\pi n}{\bar a_{+,i}}\right)$.
The standard branch-resolved observables are
\begin{equation}
 \theta_\infty^+=\frac{u_m^+}{D_L},
 \qquad
 s_i^+=\theta_\infty^+
 \exp\left(\frac{\bar b_{+,i}-2\pi}{\bar a_{+,i}}\right),
 \qquad
 r_{{\rm mag},i}^+=\frac{5\pi}{\bar a_{+,i}\ln10}.
\label{eq:combined_sdl_observables}
\end{equation}
Here, $\theta_\infty^+$ is the limiting angular position of the prograde
equatorial image sequence, not the radius or diameter of the complete
two-dimensional shadow.

\subsection{Same-side time delay}

Different relativistic images follow different photon paths and hence
need not reach the observer simultaneously.  Dividing $\dot t$ by
$\dot\rho$ in Eq.~\eqref{eq:combined_lensing_first_order} gives the change in
coordinate time along either radial part of the trajectory,
\begin{equation}
 \frac{\dd t}{\dd\rho}=
 \frac{\sqrt{\mathcal B_i}(2\mathcal C_i-\mathcal D_i u)}
 {\sqrt{4\mathcal A_i\mathcal C_i+\mathcal D_i^2}
 \sqrt{\mathcal C_i-\mathcal D_i u-\mathcal A_i u^2}}.
\label{eq:combined_time_integrand}
\end{equation}
The Bozza--Mancini treatment separates the large contribution accumulated
near the critical photon orbit from finite contributions along the rest of
the path.  Keeping the logarithmically enhanced part and the usual leading
finite correction gives the standard two-term SDL approximation for two
images of orders $n>j$ in the same sequence~\cite{Bozza:2003cp},
\begin{align}
 \Delta T_{n,j}^{s,+}\simeq{}&2\pi(n-j)
 \frac{\widetilde a_{+,i}}{\bar a_{+,i}}
\notag\\
 &+2\sqrt{\frac{\mathcal B_{im}}{\mathcal A_{im}}}
 \sqrt{\frac{u_m^+}{\bar c_i}}
 e^{\bar b_{+,i}/(2\bar a_{+,i})}
 \left[
 e^{-(2j\pi\mp\gamma)/(2\bar a_{+,i})}
 -e^{-(2n\pi\mp\gamma)/(2\bar a_{+,i})}
 \right].
\label{eq:combined_time_delay_general}
\end{align}
The upper or lower sign specifies which side of the source contains the two
images, and $\gamma\simeq0$ for the aligned configuration considered here.
The critical relation gives
\begin{equation}
 \frac{\widetilde a_{+,i}}{\bar a_{+,i}}=u_m^+.
\label{eq:combined_atilde_ratio}
\end{equation}
For the second and first images,
\begin{align}
 \Delta T_{2,1}^{+,i}
 &\simeq\underbrace{2\pi u_m^+}_{\Delta T_{\rm wind}^+}
 +\underbrace{2\sqrt{\frac{\mathcal B_{im}}{\mathcal A_{im}}}
 \sqrt{\frac{u_m^+}{\bar c_i}}
 \left[
 e^{(\bar b_{+,i}-2\pi)/(2\bar a_{+,i})}
 -e^{(\bar b_{+,i}-4\pi)/(2\bar a_{+,i})}
 \right]}_{\Delta T_{{\rm exp},i}^+}.
\label{eq:combined_time_delay_two_terms}
\end{align}
Equation~\eqref{eq:combined_time_delay_two_terms} retains the two standard
SDL terms and is not a direct integration of the complete travel-time
difference.
The compact correction in
Eq.~\eqref{eq:combined_time_delay_two_terms} contains
$\sqrt{\mathcal B_{im}/\mathcal A_{im}}$.  Since
$\mathcal A_{im}=1-2m/R_m^+$, this factor is explicitly real when the
prograde critical orbit lies outside the equatorial ergoregion.  For the
Kerr-like critical radius used here, this requires
$a/m<1/\sqrt{2}\simeq0.7071$.  We therefore show the representative values
$a/m=0,0.3,0.6$.  This restriction concerns the compact correction term, not
the existence of a physical time delay: the original coordinate-time
integrand remains real at higher spin.  Evaluating that regime requires the
full finite integrals, or an equivalent formulation in which every
intermediate factor remains real, so it is not attempted here.

\begin{figure*}[t!]
\centering
\includegraphics[width=0.98\textwidth]
{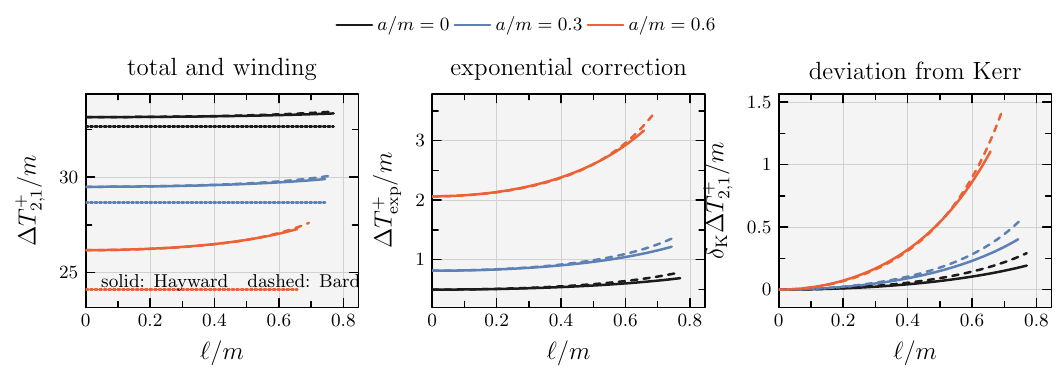}
\caption{Two-term same-side prograde SDL time-delay approximation for
$a/m=0,0.3,0.6$.
Solid and dashed curves denote the Hayward-like and Bardeen-like models;
dotted curves in the first panel show the common winding term.  All curves
are restricted to their model's black-hole domain.}
\label{fig:combined_time_delay}
\end{figure*}

\begin{figure*}[t!]
\centering
\includegraphics[width=0.86\textwidth]
{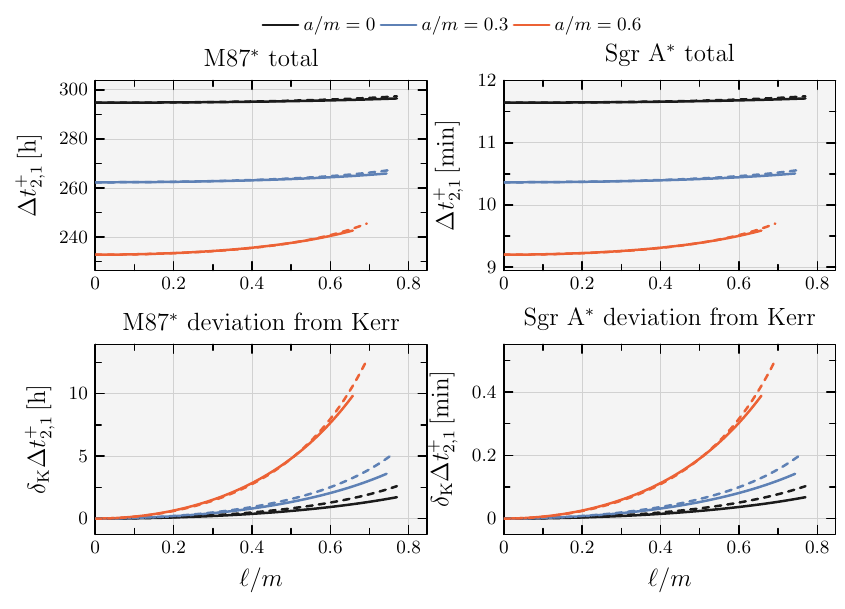}
\caption{Source-scaled two-term same-side prograde SDL time-delay approximation
and same-spin Kerr deviation for M87* and Sgr~A*. Solid and dashed curves denote the Hayward-like and Bardeen-like models. The mass conversion follows
Eq.~\eqref{eq:combined_time_conversion}.}
\label{fig:combined_time_delay_sources}
\end{figure*}

At fixed spin, the dominant winding term is common to both models and is
independent of $\ell$.  Within the two-term approximation, their difference comes from the smaller correction term.  We compare each result
with Kerr at the same spin by defining
\begin{equation}
 \delta_{\rm K}\Delta T_{2,1}^{+,i}(a,\ell)
 =\Delta T_{2,1}^{+,i}(a,\ell)-\Delta T_{2,1}^{+}(a,0).
\label{eq:combined_time_delay_kerr_deviation}
\end{equation}
Figures~\ref{fig:combined_time_delay} and
\ref{fig:combined_time_delay_sources} show both contributions and their
source-scaled values.  A dimensionless result is converted through
\begin{equation}
 \Delta t_{2,1}^{+,i}=\frac{GM}{c^3}
 \frac{\Delta T_{2,1}^{+,i}}{m}.
\label{eq:combined_time_conversion}
\end{equation}

\FloatBarrier

\subsection{Source-scaled SDL observables and comparison with Kerr}

We use $M=6.5\times10^9M_\odot$ and $D_L=16.8\,{\rm Mpc}$ for M87*
\cite{EventHorizonTelescope:2019pgp,EventHorizonTelescope:2019ggy}, and
$M=4.28\times10^6M_\odot$ and $D_L=8.32\,{\rm kpc}$ for Sgr~A*
\cite{gillessen2017update}.  The angular conversion is
\begin{equation}
 \theta_\infty^+=\frac{u_m^+}{m}
 \left(0.0098706287\,\mu{\rm as}\right)
 \frac{M/M_\odot}{D_L/{\rm pc}}.
\label{eq:combined_angular_conversion}
\end{equation}
This converts the dimensionless SDL result into an angular scale.  These
observables are not the
radius, width, or brightness contrast of the plasma emission ring reconstructed
by the EHT
\cite{EventHorizonTelescope:2019dse,EventHorizonTelescope:2019pgp,
EventHorizonTelescope:2022wkp,EventHorizonTelescope:2022apq}.
The same Kerr-referenced quantities are widely used in strong-lensing studies
of rotating and regular metrics
\cite{rotatingbh1,regularbh1,Xie:2024srr,Guo:2025rrbl,rotatingbh4}.

Table~\ref{tab:combined_sdl_observables} lists a common absolute $\ell/m$
grid.  Only black-hole entries are reported; dashes mark values outside the
corresponding outer-horizon domain.
\begin{table*}[t!]
\centering
\caption{Prograde equatorial SDL estimates.  The limiting angles are
common to both geometries, while the separations and relative magnitudes are
model dependent.  The $\ell=0$ row in each spin block is the corresponding
same-spin Kerr benchmark.}
\label{tab:combined_sdl_observables}
\tabletypesize{\scriptsize}
\setlength{\tabcolsep}{2.4pt}
\renewcommand{\arraystretch}{1.08}
\begin{tabular}{cccccccccc}
\toprule
&&\multicolumn{2}{c}{limiting angle}
&\multicolumn{2}{c}{M87* $s^+$}
&\multicolumn{2}{c}{Sgr~A* $s^+$}
&\multicolumn{2}{c}{$r_{\rm mag}^+$}\\
\cmidrule(lr){3-4}\cmidrule(lr){5-6}\cmidrule(lr){7-8}\cmidrule(lr){9-10}
$a/m$&$\ell/m$&$\theta_\infty^{\rm M87*}$&$\theta_\infty^{\rm SgrA*}$
&H&B&H&B&H&B\\
&&\multicolumn{8}{c}{angular quantities in $\mu{\rm as}$}\\
\midrule
0.0 & 0.00 & 19.8441 & 26.3844 & 0.02483 & 0.02483 & 0.03302 & 0.03302 & 6.822 & 6.822 \\
 & 0.20 & 19.8441 & 26.3844 & 0.02558 & 0.02581 & 0.03402 & 0.03432 & 6.781 & 6.776 \\
 & 0.40 & 19.8441 & 26.3844 & 0.02802 & 0.02915 & 0.03726 & 0.03876 & 6.654 & 6.627 \\
 & 0.60 & 19.8441 & 26.3844 & 0.03282 & 0.03646 & 0.04364 & 0.04848 & 6.425 & 6.338 \\
 & 0.77 & 19.8441 & 26.3844 & -- & -- & -- & -- & -- & -- \\
\midrule
0.3 & 0.00 & 17.4345 & 23.1807 & 0.04980 & 0.04980 & 0.06622 & 0.06622 & 5.938 & 5.938 \\
 & 0.20 & 17.4345 & 23.1807 & 0.05166 & 0.05197 & 0.06869 & 0.06909 & 5.885 & 5.885 \\
 & 0.40 & 17.4345 & 23.1807 & 0.05783 & 0.05949 & 0.07689 & 0.07910 & 5.717 & 5.712 \\
 & 0.60 & 17.4345 & 23.1807 & 0.07048 & 0.07672 & 0.09371 & 0.10200 & 5.401 & 5.357 \\
 & 0.77 & 17.4345 & 23.1807 & -- & -- & -- & -- & -- & -- \\
\midrule
0.6 & 0.00 & 14.6592 & 19.4906 & 0.12281 & 0.12281 & 0.16328 & 0.16328 & 4.748 & 4.748 \\
 & 0.20 & 14.6592 & 19.4906 & 0.12861 & 0.12862 & 0.17100 & 0.17102 & 4.673 & 4.686 \\
 & 0.40 & 14.6592 & 19.4906 & 0.14830 & 0.14934 & 0.19717 & 0.19856 & 4.428 & 4.475 \\
 & 0.60 & 14.6592 & 19.4906 & 0.18890 & 0.19852 & 0.25117 & 0.26395 & 3.907 & 3.983 \\
 & 0.77 & 14.6592 & 19.4906 & -- & -- & -- & -- & -- & -- \\
\midrule
0.9 & 0.00 & 10.8628 & 14.4431 & 0.50008 & 0.50008 & 0.66490 & 0.66490 & 2.641 & 2.641 \\
 & 0.20 & 10.8628 & 14.4431 & 0.52281 & 0.52022 & 0.69513 & 0.69168 & 2.521 & 2.571 \\
 & 0.40 & 10.8628 & 14.4431 & 0.55460 & 0.58364 & 0.73739 & 0.77600 & 2.015 & 2.287 \\
 & 0.60 & 10.8628 & 14.4431 & -- & -- & -- & -- & -- & -- \\
 & 0.77 & 10.8628 & 14.4431 & -- & -- & -- & -- & -- & -- \\
\bottomrule

\end{tabular}
\end{table*}

\begin{figure*}[t!]
\centering
\includegraphics[width=0.98\textwidth]
{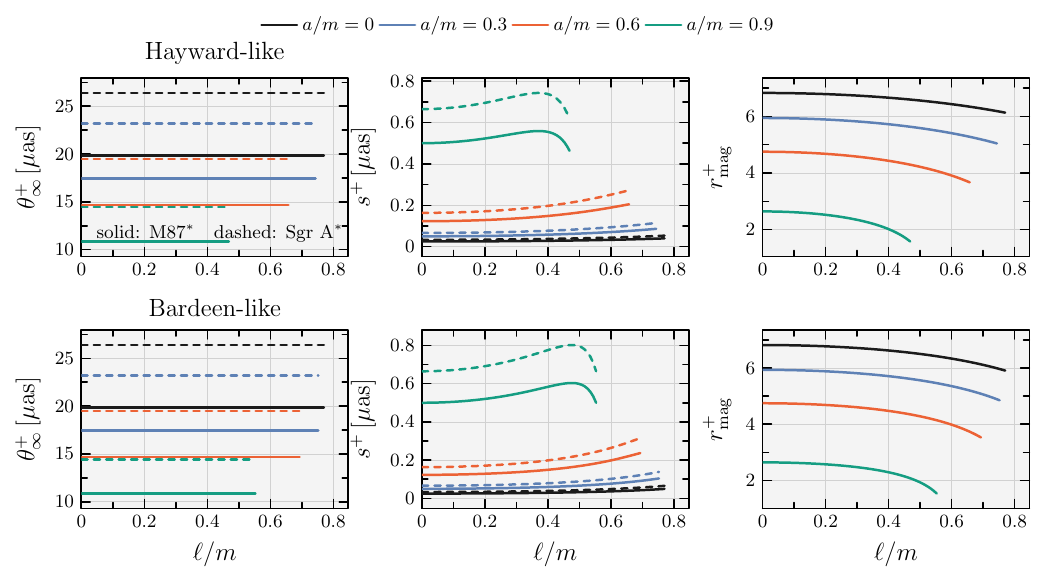}
\caption{Black-hole-only prograde image observables for
$a/m=0,0.3,0.6,0.9$.  Solid and dashed curves in the angular panels denote
M87* and Sgr~A*, respectively.  Every curve ends at the appropriate
model-dependent outer-horizon bound and begins at its same-spin Kerr value at
$\ell=0$.}
\label{fig:combined_lensing_observables}
\end{figure*}

For comparison, an $\ell>0$ row is first compared with the $\ell=0$ Kerr
value in the same spin block; Hayward-like and Bardeen-like entries are then
compared within the same row when both describe black holes.  The limiting
angle $\theta_\infty^+$ falls from $19.8441$ to $10.8628\,\mu{\rm as}$ for
M87* and from $26.3844$ to $14.4431\,\mu{\rm as}$ for Sgr~A* as $a/m$
increases from $0$ to $0.9$.  At fixed spin, $\theta_\infty^+$ is independent of $\ell$ and
common to both geometries, so it cannot distinguish the three metrics.

The separation $s^+=\theta_1^+-\theta_\infty^+$ does distinguish them.  At
$a/m=0.6$, $\ell/m=0.6$, the Kerr, Hayward-like, and Bardeen-like values are
$0.12281$, $0.18890$, and $0.19852\,\mu{\rm as}$ for M87*, and $0.16328$,
$0.25117$, and $0.26395\,\mu{\rm as}$ for Sgr~A*.  The deviation from Kerr
are $0.06609$--$0.10067\,\mu{\rm as}$, while the Hayward--Bardeen differences
are $0.00962$--$0.01278\,\mu{\rm as}$.  Hence, nominal $1\sigma$ sensitivity
of order $0.1\,\mu{\rm as}$ ($100$ nanoarcseconds) can begin testing a regular geometry against
Kerr, whereas distinguishing all three requires about $0.01\,\mu{\rm as}$
($10$ nanoarcseconds).

The relative magnitude $r_{{\rm mag},i}^+$ measures how strongly the
higher-order images are suppressed.  Its corresponding flux ratio is
\begin{equation}
 \mathcal F_i^+\equiv
 \frac{\mu_{1,i}^+}{\displaystyle\sum_{n=2}^{\infty}\mu_{n,i}^+}
 \simeq10^{0.4r_{{\rm mag},i}^+}
 =\exp\left(\frac{2\pi}{\bar a_{+,i}}\right),
\label{eq:combined_sdl_flux_ratio}
\end{equation}
where $\mu_{n,i}^+$ is the absolute magnification of the $n$th prograde image.
This is the conventional leading SDL estimate.  Retaining the complete
geometric image tail changes the exponential factor to
$\exp(2\pi/\bar a_{+,i})-1$ at the same order.
At the same parameter point, $r_{\rm mag}^+$ is $4.748$, $3.907$, and $3.983$, respectively for Kerr, Hayward-like, and Bardeen-like, corresponding to flux ratios
$79.3$, $36.5$, and $39.2$.  The regular metrics, therefore, make the
higher-order images relatively brighter than Kerr.  Their differences from
Kerr are $0.765$--$0.841$ mag, while the Hayward--Bardeen difference is
$0.076$ mag.  Nominal photometric precision of order $0.8$ mag can begin
testing a regular geometry against Kerr, whereas about $0.08$ mag, equivalent
to roughly $7\%$ precision in the flux ratio, is needed to distinguish all
three.

The time delay gives a third discriminator.  At the same point, the Kerr,
Hayward-like, and Bardeen-like delays are respectively $232.854$, $240.512$, and
$240.833\,{\rm h}$ for M87*, and $9.1995$, $9.5021$, and
$9.5148\,{\rm min}$ for Sgr~A*.  Relative to Kerr, the shifts are
$7.659$--$7.980\,{\rm h}$ and $18.15$--$18.92\,{\rm s}$; the closer
Hayward--Bardeen differences are $19.3\,{\rm min}$ and $0.761\,{\rm s}$.
Within the two-term SDL approximation, the Kerr--regular shifts amount to
several hours for M87* and tens of seconds for Sgr~A*.  This suggests that in principle, time-delay measurements from a suitable, strongly
lensed source could help distinguish either regular geometry from Kerr.  The
smaller Hayward--Bardeen difference would be more difficult to resolve.

Current EHT observations do not directly measure $s^+$,
$r_{\rm mag}^+$, or the time delay, because these quantities require the
individual relativistic-image contributions to be separated
\cite{EventHorizonTelescope:2019dse,EventHorizonTelescope:2022apq}.  In the following subsection, we compare our calculated full shadow with the EHT
observations.

\subsection{Kerr-degenerate shadow and EHT scale comparison}

The complete rotating shadow is determined from the full celestial curve
$(X(R_p),Y(R_p))$, whereas $\theta_\infty^+$ describes only the prograde
equatorial image sequence.  Following standard shadow analyses, we
characterize its overall size by the area-equivalent radius $R_{\rm A}$, the
radius of a circle with the same area.  This is a convenient summary of the
complete boundary rather than a unique definition of the shadow
\cite{Kumar:2020shadowParameters,Afrin:2021hairyKerr,Ali:2024rqcbh}.
If $R_{p,1}$ and $R_{p,2}$ are endpoints of $R_{\rm A}$ at $Y=0$, then
\begin{equation}
 \mathcal A_{\rm sh}=2\left|\int^{R_{p,2}}_{R_{p,1}} Y(R_p)
 \frac{\dd X(R_p)}{\dd R_p}\dd R_p\right|,
 \qquad
 R_{\rm A}=\sqrt{\frac{\mathcal A_{\rm sh}}{\pi}},
\label{eq:combined_shadow_area}
\end{equation}
and
\begin{equation}
 \delta_{\rm sh}=\frac{R_{\rm A}}{3\sqrt3m}-1,
 \qquad
 \theta_{\rm sh}=2\frac{R_{\rm A}}{m}
 \left(0.0098706287\,\mu{\rm as}\right)
 \frac{M/M_\odot}{D_L/{\rm pc}}.
\label{eq:combined_shadow_observables}
\end{equation}
Here, $\delta_{\rm sh}$ measures the fractional change relative to the
Schwarzschild critical radius $3\sqrt3m$; it is not a deviation from Kerr at
the same spin.  Thus, $R_{\rm A}$ measures the complete shadow without assuming
that it is circular.  Complementary choices in the literature include the
Hioki--Maeda reference radius and distortion, circularity, oblateness, and
axis-ratio measures
\cite{Hioki:2009shadow,Bambi:2019M87,Afrin:2021hairyKerr,rotatingbh1,
Ali:2024rqcbh,Nengroo:2026kalbRamond}.  In fact, the critical impact
parameters expressed through
$R_p$ have the Kerr form, so for fixed $(a,\theta_o)$, the entire celestial
curve obeys
\begin{equation}
 \bigl(X_i(R_p),Y_i(R_p)\bigr)
 =\bigl(X_{\rm K}(R_p),Y_{\rm K}(R_p)\bigr),
 \qquad
 R_{{\rm A},i}=R_{{\rm A},{\rm K}},
 \qquad
 \delta_{\rm K}\theta_{{\rm sh},i}
 \equiv\theta_{{\rm sh},i}(a,\ell)-\theta_{{\rm sh},{\rm K}}(a)=0,
\label{eq:combined_shadow_kerr_degeneracy}
\end{equation}
provided that the required exterior spherical-photon interval belongs to the
physical domain.  Thus, $\ell$ changes the coordinate photon radii and the
admissible black-hole parameter domain but not the shadow projected on the
observer's screen.  At fixed $(m,a,\theta_o)$, the shadow area, displacement,
and shape are identical to Kerr and independent of $\ell$.  Shadow-only
observables therefore cannot determine $\ell$ in this ansatz; the model
dependence retained in Table~\ref{tab:combined_eht_shadow} comes only from the
different admissible parameter bounds.

The observer inclination $\theta_o$ is the angle between the line of sight
and the assumed spin axis.  For M87*, we use $\theta_o=17^\circ$, motivated by
the measured orientation of the large-scale jet and the usual assumption
that the spin and jet axes are aligned
\cite{Walker:2018mrv,EventHorizonTelescope:2019pgp}.  The same orientation is
sometimes quoted as $163^\circ$ because the spin axis is a directed vector;
the ideal shadow is unchanged under
$\theta_o\rightarrow180^\circ-\theta_o$.  For Sgr~A*, the spin orientation is
not well determined.  We use $\theta_o=50^\circ$ as a relatively inclined
benchmark because the EHT model comparisons tend to disfavor inclinations
larger than about $50^\circ$; it should not be read as a measured inclination
\cite{EventHorizonTelescope:2022urf,EventHorizonTelescope:2022xqj}.

For M87*, the directly reconstructed quantity is the bright emission ring
diameter $d_{\rm ring}=42\pm3\,\mu{\rm as}$, not the geometric critical-curve
diameter calculated here
\cite{EventHorizonTelescope:2019dse,EventHorizonTelescope:2019ggy}.  Some
phenomenological shadow studies use the simple estimate

\begin{equation}
 d_{\rm sh}\simeq0.9d_{\rm ring}
 =37.8\pm2.7\,\mu{\rm as},
\label{eq:combined_m87_ring_proxy}
\end{equation}
or $35.1$--$40.5\,\mu{\rm as}$ at the quoted $1\sigma$ level
\cite{Banerjee:2022bardeenEHT,Ali:2024rqcbh}.  The roughly $10\%$ reduction is
intended to represent the offset between the peak of a plasma-emission ring
and the underlying lensing critical curve. The offset
depends on the emission and calibration model, and a $10\%$ reduction is not a universal EHT conversion.  We shall quote
Eq.~\eqref{eq:combined_m87_ring_proxy} only as an illustrative scale proxy.
EHT Paper~VI also reported
$\delta_g=-0.01\pm0.17$, where $\delta_g$ compares the gravitational angular
scale inferred from the EHT image with the stellar-dynamical prior; it is not
defined identically to our $\delta_{\rm sh}$
\cite{EventHorizonTelescope:2019ggy}.  Following the translation used in
shadow tests~\cite{Kocherlakota:2021dcv,Afrin:2024khy}, we use the corresponding interval
only as an approximate scale-level consistency check.

For Sgr~A*, the EHT image shows an angular shadow diameter $d_{\rm sh}=48.7\pm7.0\,\mu{\rm as}$
\cite{EventHorizonTelescope:2022urf,EventHorizonTelescope:2022wkp,
EventHorizonTelescope:2022xqj}.  The corresponding fractional bounds are
$\delta_{\rm SgrA*}=-0.08\pm0.09$ with the VLTI mass--distance prior and
$-0.04^{+0.09}_{-0.10}$ with the Keck prior
\cite{EventHorizonTelescope:2022xqj}.  Unlike the fixed $10\%$ M87* proxy,
these Sgr~A* quantities come from an EHT calibration of the relation between
the emission ring and the geometric shadow using image simulations.  We use
the quoted bounds only as consistency checks and do not construct a new EHT
likelihood.

\begin{table}[h!]
\centering
\caption{Area-equivalent geometric-shadow estimates.  Each row equals the
same-spin Kerr prediction and is unchanged across values of $\ell$ below the
corresponding Hayward-like or Bardeen-like black-hole bound.  The quantity
$\delta_{\rm sh}$ is measured relative to Schwarzschild, not to Kerr at the
same spin.  The final row gives the quoted $1\sigma$ comparison intervals.
The M87* angular interval uses the illustrative $10\%$ ring reduction; the
Sgr~A* fractional interval uses the VLTI prior.}
\label{tab:combined_eht_shadow}
\tabletypesize{\scriptsize}
\setlength{\tabcolsep}{3.0pt}
\renewcommand{\arraystretch}{1.10}
\begin{tabular}{ccccccc}
\toprule
$a/m$&$\ell_{h,H}^+/m$&$\ell_{h,B}^+/m$
&$\theta_{\rm sh}^{\rm M87*}$&$\delta_{\rm sh}^{\rm M87*}$
&$\theta_{\rm sh}^{\rm SgrA*}$&$\delta_{\rm sh}^{\rm SgrA*}$\\
\midrule
0.0&0.769800&0.769800&39.6881&0&52.7688&0\\
0.3&0.743361&0.752072&39.4950&-0.004867&52.5786&-0.003605\\
0.6&0.657267&0.692820&38.8715&-0.020577&51.9529&-0.015462\\
0.9&0.468290&0.552674&37.6132&-0.052280&50.5766&-0.041543\\
\midrule
\multicolumn{3}{l}{Adopted approximate $1\sigma$ comparison intervals}
&35.1--40.5&$[-0.18,0.16]$&41.7--55.7&$[-0.17,0.01]$\\
\bottomrule
\end{tabular}
\end{table}

Every theoretical value in Table~\ref{tab:combined_eht_shadow} lies inside
the corresponding approximate comparison interval in the final row.  The
Sgr~A* values also lie inside the Keck fractional interval
$-0.14\leq\delta_{\rm SgrA*}\leq0.05$.  Thus, at the level of these shadow
size bounds, the displayed rotating Hayward-like and Bardeen-like black holes
are consistent with the current M87* and Sgr~A* observations.  This consistency does not favor either regular geometry over Kerr, select a preferred spin, or determine the value of the regularity length $\ell$.

The variation across the rows is entirely a spin effect relative to
Schwarzschild.  For example, at $a/m=0.9$ the M87* diameter is
$37.6132\,\mu{\rm as}$ with $\delta_{\rm sh}=-0.0523$, while the Sgr~A*
diameter is $50.5766\,\mu{\rm as}$ with
$\delta_{\rm sh}=-0.0415$.

Indeed, the Hayward-like, Bardeen-like, and Kerr shadows are identical at the
same spin and inclination on their common black-hole domain.  Their only
difference in Table~\ref{tab:combined_eht_shadow} is the allowed range of
$\ell/m$: For $a>0$, the Bardeen-like geometry admits a somewhat larger
outer-horizon domain.

Some equatorial SDL studies use $2\theta_\infty$ as a proxy for the rotating
shadow diameter~\cite{rotatingbh2,loop1}.  In the present work,
$\theta_\infty^+$ is only the limiting position of the prograde equatorial
image sequence.  We do not identify $2\theta_\infty^+$ with the
full rotating shadow; instead, the EHT comparison uses the area-equivalent
diameter calculated from the complete celestial curve.

  The reported agreement with the quoted fractional bounds is a consistency check, not evidence for either metric, and puts no constraint on \(\ell\).
Moreover, the exact same-spin Kerr degeneracy in
Eq.~\eqref{eq:combined_shadow_kerr_degeneracy} prevents the ideal shadow
boundary from distinguishing the two geometries in the first place.  Within
the present construction, such a distinction must instead come from
radial-measure-sensitive quantities, such as the SDL separations or time
delays studied above.  Connecting those predictions to interferometric data
would require a source model, radiative transfer, and a statistical treatment
of observational and calibration uncertainties.
\FloatBarrier

\section{Discussion and Conclusions}
\label{sec:combined_conclusions}

We have compared rotating Hayward-like and Bardeen-like geometries obtained
from two areal-radius functions within the same Kerr-like ansatz.  This setup
separates consequences of their common Kerr dependence from effects of their
different radial profiles.  The conclusions below apply to this construction
and to the adopted radial branches, rather than to all rotating regular black
holes.

On each branch, the positive minimum of the areal radius excludes the Kerr
ring locus $\Sigma_i=0$, and the curvature invariants examined here remain
finite.  This supports scalar-curvature regularity on the stated domain but
does not establish geodesic completeness through its endpoint.  Depending on
$(a,\ell)$, the branch contains two horizons, only the outer horizon, or no
horizon; the Bardeen-like outer-horizon domain is somewhat larger at nonzero
spin.  The status and stability of an inner horizon in the two-horizon region
would require a separate global analysis.  The finite-surface Komar charges
approach $M_K\to m$ and $J_K\to am$ asymptotically.  When the Einstein tensor
is interpreted as an effective stress tensor, the negative equatorial radial
null contraction found for every $\ell>0$ is sufficient to violate the NEC
and WEC.  This is a limitation of the present effective geometry, not a
general no-go result for rotating regular spacetimes.

The null Hamilton--Jacobi equation separates in both models.  Expressed in
terms of the areal photon radius, their critical impact parameters take the
Kerr form.  Hence, whenever the complete spherical-photon family belongs to
the physical branch, the ideal shadow is identical to Kerr at the same
$(m,a,\theta_o)$.  As such, the shadow's area, displacement, and shape, therefore, cannot
constrain $\ell$ or distinguish the Hayward-like and Bardeen-like geometries.
The M87* and Sgr~A* comparisons are consequently compatibility checks against
published EHT-based bounds.  The $10\%$ reduction of the M87* emission-ring diameter is treated as a model-dependent literature proxy.

The strong-lensing observables retain model dependence.  Although
$u_m^+$ and $\theta_\infty^+$ are common at fixed spin, $\bar a_+$ contains
$1/R_i'(\rho_{m,i})$ and $\bar b_+$ depends on the full regularized integral.
The predicted image separations, leading SDL flux ratios, and two-term
time-delay estimates are different.  At $a/m=0.6$ and
$\ell/m=0.6$, nominal sensitivities of
order $0.1\,\mu{\rm as}$ could test a regular geometry against Kerr, while
about $0.01\,\mu{\rm as}$ would be needed to separate all three predictions.
The corresponding illustrative photometric scales are approximately $0.8$
and $0.08$ mag.  For M87* (Sgr~A*), the two-term time-delay approximation
separates the curves by about $7.7\,{\rm h}$ ($18\,{\rm s}$) relative to Kerr
and $19\,{\rm min}$ ($0.8\,{\rm s}$) between the two regular geometries.
For a suitable, strongly lensed source, time
delays could in principle help distinguish the regular geometries from Kerr,
although the smaller Hayward--Bardeen separation would be more demanding.
These values remain predictions of the compact SDL approximation rather than
instrumental resolution requirements.

The main result is that at fixed mass, spin, and observer inclination, the
ideal shadows of the Kerr, Hayward-like, and Bardeen-like geometries are the
same.  By contrast, their image separations, relative brightnesses, and
two-term time-delay estimates can differ because these quantities depend
on the radial motion of the photons.  Determining whether such differences
can be measured will require more realistic calculations that include photon
motion away from the equatorial plane, the source and plasma emission, and
instrumental and astrophysical uncertainties.  It also remains to identify a
physical matter source for the rotating metrics, determine whether the
spacetime can be continued through the minimum-radius boundary, and study the
stability of any inner horizon.

\section*{Acknowledgements}
 YW is supported by NSFC Grant No. 12475001, the Shanghai Municipal Science and Technology Major Project (Grant No. 2019SHZDZX01), Science and Technology Commission of Shanghai Municipality (Grant No. 24LZ1400100), and the Innovation Program for Quantum Science and Technology (No. 2024ZD0300101).

\clearpage
\appendix
\section{Hamilton--Jacobi Separation}
\label{app:combined_hj_separation}

Stationarity and axisymmetry alone do not guarantee a Carter quantity.  For
the present metrics, separability follows directly from the inverse metric.
With
\begin{equation}
 B_i(\rho)=R_i^2(\rho)+a^2,
\end{equation}
its nonzero components satisfy
\begin{align}
 \Sigma_i g^{tt}&=-\frac{B_i^2}{\Delta_i}
 +a^2\sin^2\theta,\notag\\
 \Sigma_i g^{t\phi}&=-\frac{a(B_i-\Delta_i)}{\Delta_i},
 \qquad
 \Sigma_i g^{\phi\phi}=\csc^2\theta-\frac{a^2}{\Delta_i},
 \notag\\
 \Sigma_i g^{\rho\rho}&=\Delta_i,
 \qquad
 \Sigma_i g^{\theta\theta}=1.
\label{eq:combined_inverse_metric}
\end{align}
Each component of $\Sigma_i g^{\mu\nu}$ separates into radial and angular
parts.  This is the Carter--Staeckel structure relevant to the null
Hamilton--Jacobi equation
\cite{Carter:1968separable,Benenti:1979separability,
Chandrasekhar:1983blackholes}.

Insert the action
\begin{equation}
 S=-Et+L_z\phi+S_\rho(\rho)+S_\theta(\theta)
\label{eq:combined_hj_ansatz}
\end{equation}
into $g^{\mu\nu}\partial_\mu S\partial_\nu S=0$.  Using
Eq.~\eqref{eq:combined_inverse_metric} and multiplying by $\Sigma_i$ gives
\begin{align}
0={}&\left[
 \Delta_i(S_\rho')^2-
 \frac{(B_iE-aL_z)^2}{\Delta_i}
 \right]
\notag\\
&+\left[
 (S_\theta')^2+
 (L_z\csc\theta-aE\sin\theta)^2
 \right].
\label{eq:combined_hj_separated_sum}
\end{align}
The first bracket depends only on $\rho$ and the second only on $\theta$, so
they equal opposite constants.  Writing the separation constant as
$\mathcal K_{\rm sep}$ gives
\begin{align}
 \Delta_i(S_\rho')^2-
 \frac{(B_iE-aL_z)^2}{\Delta_i}
 &=-\mathcal K_{\rm sep},
 \notag\\
 (S_\theta')^2+
 (L_z\csc\theta-aE\sin\theta)^2
 &=\mathcal K_{\rm sep}.
\end{align}
We use the standard Kerr convention
\begin{equation}
 \mathcal Q\equiv\mathcal K_{\rm sep}-(L_z-aE)^2.
\label{eq:combined_carter_convention}
\end{equation}
The separated equations can then be written as
\begin{align}
 \mathcal R_i^{(E)}(\rho)
 &=(B_iE-aL_z)^2
 -\Delta_i\left[\mathcal Q+(L_z-aE)^2\right],
 \label{eq:combined_dimful_radial_potential}\\
 \Theta(\theta)
 &=\mathcal Q-
 \left(L_z^2\csc^2\theta-a^2E^2\right)\cos^2\theta,
\end{align}
with
\begin{equation}
 \left(\Sigma_i\frac{d\rho}{d\lambda}\right)^2
 =\mathcal R_i^{(E)}(\rho),
 \qquad
 \left(\Sigma_i\frac{d\theta}{d\lambda}\right)^2
 =\Theta(\theta).
\end{equation}
Dividing by $E^2$ and introducing $\xi=L_z/E$ and
$\eta=\mathcal Q/E^2$ gives the potentials used in
Eqs.~\eqref{eq:combined_radial_potential} and
\eqref{eq:combined_angular_potential}.

For completeness, the first two spherical-orbit conditions in
Eq.~\eqref{eq:combined_spherical_conditions} can be solved by defining
$\mathcal U=\eta+(a-\xi)^2$.  At fixed conserved quantities they become
\begin{align}
 [B_i-a\xi]^2&=\Delta_i\mathcal U,
 \label{eq:combined_double_root_1}\\
 2B_i'[B_i-a\xi]&=\Delta_i'\mathcal U.
 \label{eq:combined_double_root_2}
\end{align}
For the nonprincipal branch relevant to the shadow,
\begin{equation}
 B_i-a\xi=\frac{2B_i'\Delta_i}{\Delta_i'},
 \qquad
 \mathcal U=\frac{4(B_i')^2\Delta_i}{(\Delta_i')^2}.
\end{equation}
Substitution gives Eqs.~\eqref{eq:combined_xi_general} and
\eqref{eq:combined_eta_general} in the main text.

\section{From a Null-Energy Violation to a Weak-Energy Violation}
\label{app:nec_implies_wec}

At a point where the local Carter frame is defined, consider an observer
moving radially with speed $v<1$,
\begin{equation}
 u^{(A)}=\gamma(1,v,0,0),
 \qquad \gamma=(1-v^2)^{-1/2}.
\end{equation}
The measured energy density is
\begin{equation}
 \mathcal E_{\mathrm{obs}}(v)
 =T_{(A)(B)}u^{(A)}u^{(B)}
 =\gamma^2\left[T_{(0)(0)}+2vT_{(0)(1)}
 +v^2T_{(1)(1)}\right].
\end{equation}
On the equatorial plane used in Eq.~\eqref{eq:combined_universal_nec},
$T_{(0)(1)}=0$.  If $T_{(0)(0)}<0$, the observer at rest already violates
the WEC.  Otherwise, the radial NEC violation
$T_{(0)(0)}+T_{(1)(1)}<0$ implies
$T_{(1)(1)}<-T_{(0)(0)}$ and hence
\begin{equation}
 v_*^2=\frac{T_{(0)(0)}}{-T_{(1)(1)}}<1.
\end{equation}
Any speed $v_*<v<1$ then gives
$\mathcal E_{\mathrm{obs}}(v)<0$, since $\gamma^2>0$.  Thus, a negative null
contraction is sufficient to identify at least one timelike observer who
measures negative energy density, and therefore to disprove the WEC at that
point.  This argument does not require diagonalizing $T_{(A)(B)}$.

\bibliography{QII}{}

@article{Bozza:2008ev,
    author = "Bozza, V.",
    title = "{A Comparison of approximate gravitational lens equations and a proposal for an improved new one}",
    eprint = "0807.3872",
    archivePrefix = "arXiv",
    primaryClass = "gr-qc",
    doi = "10.1103/PhysRevD.78.103005",
    journal = "Phys. Rev. D",
    volume = "78",
    pages = "103005",
    year = "2008"
}

@article{Calza:2025mrt,
    author = "Calz{\'a}, Marco and Rinaldi, Massimiliano and Zerbini, Sergio",
    title = "{Topological regular black holes without a Cauchy horizon}",
    eprint = "2505.04427",
    archivePrefix = "arXiv",
    primaryClass = "gr-qc",
    doi = "10.1103/dyy8-qjnd",
    journal = "Phys. Rev. D",
    volume = "112",
    number = "2",
    pages = "024024",
    year = "2025"
}

@article{Poisson:1990eh,
    author = "Poisson, Eric and Israel, W.",
    title = "{Internal structure of black holes}",
    doi = "10.1103/PhysRevD.41.1796",
    journal = "Phys. Rev. D",
    volume = "41",
    pages = "1796--1809",
    year = "1990"
}

@article{Bozza:2001xd,
    author = "Bozza, V. and Capozziello, S. and Iovane, G. and Scarpetta, G.",
    title = "{Strong field limit of black hole gravitational lensing}",
    eprint = "gr-qc/0102068",
    archivePrefix = "arXiv",
    doi = "10.1023/A:1012292927358",
    journal = "Gen. Rel. Grav.",
    volume = "33",
    pages = "1535--1548",
    year = "2001"
}

@article{Bozza:2002zj,
    author = "Bozza, V.",
    title = "{Gravitational lensing in the strong field limit}",
    eprint = "gr-qc/0208075",
    archivePrefix = "arXiv",
    doi = "10.1103/PhysRevD.66.103001",
    journal = "Phys. Rev. D",
    volume = "66",
    pages = "103001",
    year = "2002"
}

@article{Bozza:2003quasi,
    author = "Bozza, V.",
    title = "{Quasiequatorial gravitational lensing by spinning black holes in the strong field limit}",
    eprint = "gr-qc/0210109",
    archivePrefix = "arXiv",
    doi = "10.1103/PhysRevD.67.103006",
    journal = "Phys. Rev. D",
    volume = "67",
    pages = "103006",
    year = "2003"
}

@article{EventHorizonTelescope:2019ggy,
    author = "Akiyama, Kazunori and others",
    collaboration = "Event Horizon Telescope",
    title = "{First M87 Event Horizon Telescope Results. VI. The Shadow and Mass of the Central Black Hole}",
    eprint = "1906.11243",
    archivePrefix = "arXiv",
    primaryClass = "astro-ph.GA",
    doi = "10.3847/2041-8213/ab1141",
    journal = "Astrophys. J. Lett.",
    volume = "875",
    number = "1",
    pages = "L6",
    year = "2019"
}

@article{gillessen2017update,
  title={An update on monitoring stellar orbits in the galactic center},
  author={Gillessen, Stefan and Plewa, PM and Eisenhauer, Frank and Sari, Re'em and Waisberg, Idel and Habibi, Maryam and Pfuhl, Oliver and George, Elizabeth and Dexter, Jason and von Fellenberg, Sebastiano and others},
  journal={The Astrophysical Journal},
  volume={837},
  number={1},
  pages={30},
  year={2017},
  publisher={IOP Publishing}
}

@article{EventHorizonTelescope:2022wkp,
    author = "Akiyama, Kazunori and others",
    collaboration = "Event Horizon Telescope",
    title = "{First Sagittarius A* Event Horizon Telescope Results. I. The Shadow of the Supermassive Black Hole in the Center of the Milky Way}",
    eprint = "2311.08680",
    archivePrefix = "arXiv",
    primaryClass = "astro-ph.HE",
    doi = "10.3847/2041-8213/ac6674",
    journal = "Astrophys. J. Lett.",
    volume = "930",
    number = "2",
    pages = "L12",
    year = "2022"
}

@article{EventHorizonTelescope:2022apq,
    author = "Akiyama, Kazunori and others",
    collaboration = "Event Horizon Telescope",
    title = "{First Sagittarius A* Event Horizon Telescope Results. II. EHT and Multiwavelength Observations, Data Processing, and Calibration}",
    eprint = "2311.08679",
    archivePrefix = "arXiv",
    primaryClass = "astro-ph.HE",
    reportNumber = "FERMILAB-PUB-22-418-PPD",
    doi = "10.3847/2041-8213/ac6675",
    journal = "Astrophys. J. Lett.",
    volume = "930",
    number = "2",
    pages = "L13",
    year = "2022"
}

@article{EventHorizonTelescope:2019pgp,
    author = "Akiyama, Kazunori and others",
    collaboration = "Event Horizon Telescope",
    title = "{First M87 Event Horizon Telescope Results. V. Physical Origin of the Asymmetric Ring}",
    eprint = "1906.11242",
    archivePrefix = "arXiv",
    primaryClass = "astro-ph.GA",
    doi = "10.3847/2041-8213/ab0f43",
    journal = "Astrophys. J. Lett.",
    volume = "875",
    number = "1",
    pages = "L5",
    year = "2019"
}

@article{Zhao:2017cwk,
    author = "Zhao, Shan-Shan and Xie, Yi",
    title = "{Strong deflection gravitational lensing by a modified Hayward black hole}",
    eprint = "1704.02434",
    archivePrefix = "arXiv",
    primaryClass = "gr-qc",
    doi = "10.1140/epjc/s10052-017-4850-5",
    journal = "Eur. Phys. J. C",
    volume = "77",
    number = "5",
    pages = "272",
    year = "2017"
}

@article{Gibbons:2008rj,
    author = "Gibbons, G. W. and Werner, M. C.",
    title = "{Applications of the Gauss-Bonnet theorem to gravitational lensing}",
    eprint = "0807.0854",
    archivePrefix = "arXiv",
    primaryClass = "gr-qc",
    doi = "10.1088/0264-9381/25/23/235009",
    journal = "Class. Quant. Grav.",
    volume = "25",
    pages = "235009",
    year = "2008"
}

@article{Eiroa:2010wm,
    author = "Eiroa, Ernesto F. and Sendra, Carlos M.",
    title = "{Gravitational lensing by a regular black hole}",
    eprint = "1011.2455",
    archivePrefix = "arXiv",
    primaryClass = "gr-qc",
    doi = "10.1088/0264-9381/28/8/085008",
    journal = "Class. Quant. Grav.",
    volume = "28",
    pages = "085008",
    year = "2011"
}

@article{EventHorizonTelescope:2019dse,
    author = "Akiyama, Kazunori and others",
    collaboration = "Event Horizon Telescope",
    title = "{First M87 Event Horizon Telescope Results. I. The Shadow of the Supermassive Black Hole}",
    eprint = "1906.11238",
    archivePrefix = "arXiv",
    primaryClass = "astro-ph.GA",
    doi = "10.3847/2041-8213/ab0ec7",
    journal = "Astrophys. J. Lett.",
    volume = "875",
    pages = "L1",
    year = "2019"
}

@article{EventHorizonTelescope:2022urf,
    author = "Akiyama, Kazunori and others",
    collaboration = "Event Horizon Telescope",
    title = "{First Sagittarius A* Event Horizon Telescope Results. V. Testing Astrophysical Models of the Galactic Center Black Hole}",
    eprint = "2311.09478",
    archivePrefix = "arXiv",
    primaryClass = "astro-ph.HE",
    reportNumber = "FERMILAB-PUB-22-419-PPD",
    doi = "10.3847/2041-8213/ac6672",
    journal = "Astrophys. J. Lett.",
    volume = "930",
    number = "2",
    pages = "L16",
    year = "2022"
}

@article{EventHorizonTelescope:2022xqj,
    author = "Akiyama, Kazunori and others",
    collaboration = "Event Horizon Telescope",
    title = "{First Sagittarius A* Event Horizon Telescope Results. VI. Testing the Black Hole Metric}",
    eprint = "2311.09484",
    archivePrefix = "arXiv",
    primaryClass = "astro-ph.HE",
    reportNumber = "FERMILAB-PUB-22-422-PPD",
    doi = "10.3847/2041-8213/ac6756",
    journal = "Astrophys. J. Lett.",
    volume = "930",
    number = "2",
    pages = "L17",
    year = "2022"
}

@article{Walker:2018mrv,
    author = "Walker, R. Craig and Hardee, Phillip E. and Davies, Frederick B. and Ly, Chun and Junor, William",
    title = "{The Structure and Dynamics of the Subparsec Scale Jet in M87 Based on 50 VLBA Observations over 17 Years at 43 GHz}",
    eprint = "1802.06166",
    archivePrefix = "arXiv",
    primaryClass = "astro-ph.HE",
    doi = "10.3847/1538-4357/aaafcc",
    journal = "Astrophys. J.",
    volume = "855",
    number = "2",
    pages = "128",
    year = "2018"
}

@article{Kocherlakota:2021dcv,
    author = "Kocherlakota, Prashant and others",
    collaboration = "Event Horizon Telescope",
    title = "{Constraints on Black-Hole Charges with the 2017 EHT Observations of M87*}",
    eprint = "2105.09343",
    archivePrefix = "arXiv",
    primaryClass = "gr-qc",
    doi = "10.1103/PhysRevD.103.104047",
    journal = "Phys. Rev. D",
    volume = "103",
    number = "10",
    pages = "104047",
    year = "2021"
}

@article{Banerjee:2022bardeenEHT,
    author = "Banerjee, Indrani and Sau, Subhadip and SenGupta, Soumitra",
    title = "{Do Shadows of Sgr A* and M87* Indicate Black Holes with a Magnetic Monopole Charge?}",
    eprint = "2207.06034",
    archivePrefix = "arXiv",
    primaryClass = "gr-qc",
    journal = "arXiv e-prints",
    pages = "arXiv:2207.06034",
    year = "2022"
}

@article{Bozza:2003cp,
    author = "Bozza, V. and Mancini, L.",
    title = "{Time delay in black hole gravitational lensing as a distance estimator}",
    eprint = "gr-qc/0305007",
    archivePrefix = "arXiv",
    doi = "10.1023/B:GERG.0000010486.58026.4f",
    journal = "Gen. Rel. Grav.",
    volume = "36",
    pages = "435--450",
    year = "2004"
}

@article{Bartelmann:2016dvf,
    author = "Bartelmann, Matthias and Maturi, Matteo",
    title = "{Weak gravitational lensing}",
    eprint = "1612.06535",
    archivePrefix = "arXiv",
    primaryClass = "astro-ph.CO",
    doi = "10.4249/scholarpedia.32440",
    journal = "Scholarpedia",
    volume = "12",
    number = "1",
    pages = "32440",
    year = "2017"
}

@article{Hayward:2005gi,
    author = "Hayward, Sean A.",
    title = "{Formation and evaporation of regular black holes}",
    eprint = "gr-qc/0506126",
    archivePrefix = "arXiv",
    doi = "10.1103/PhysRevLett.96.031103",
    journal = "Phys. Rev. Lett.",
    volume = "96",
    pages = "031103",
    year = "2006"
}

@INPROCEEDINGS{1968qtr..conf...87B,
       author = {{Bardeen}, James},
        title = "{Non-singular general relativistic gravitational collapse}",
    booktitle = {Proceedings of the 5th International Conference on Gravitation and the Theory of Relativity},
         year = 1968,
        month = sep,
        pages = {87},
       url = {https://ui.adsabs.harvard.edu/abs/1968qtr..conf...87B}
}

@article{Maeda:2005yd,
    author = "Maeda, Hideki and Torii, Takashi and Harada, Tomohiro",
    title = "{Novel Cauchy-horizon instability}",
    eprint = "gr-qc/0501042",
    archivePrefix = "arXiv",
    reportNumber = "WU-AP-208-05",
    doi = "10.1103/PhysRevD.71.064015",
    journal = "Phys. Rev. D",
    volume = "71",
    pages = "064015",
    year = "2005"
}

@article{Wambsganss:1998gg,
    author        = {Wambsganss, Joachim},
    title         = {{Gravitational lensing in astronomy}},
    eprint        = {astro-ph/9812021},
    archiveprefix = {arXiv},
    doi           = {10.12942/lrr-1998-12},
    journal       = {Living Rev. Rel.},
    volume        = {1},
    pages         = {12},
    year          = {1998},
}

@article{Virbhadra:1999nm,
    author        = {Virbhadra, K. S. and Ellis, George F. R.},
    title         = {Schwarzschild black hole lensing},
    eprint        = {astro-ph/9904193},
    archiveprefix = {arXiv},
    doi           = {10.1103/PhysRevD.62.084003},
    journal       = {Phys. Rev. D},
    volume        = {62},
    pages         = {084003},
    year          = {2000},
}

@article{rotatingbh1,
   title={Parameters estimation and strong gravitational lensing of nonsingular Kerr-Sen black holes},
   volume={2021},
   ISSN={1475-7516},
   url={http://dx.doi.org/10.1088/1475-7516/2021/03/056},
   DOI={10.1088/1475-7516/2021/03/056},
   number={03},
   journal={Journal of Cosmology and Astroparticle Physics},
   publisher={IOP Publishing},
   author={Ghosh, Sushant G. and Kumar, Rahul and Islam, Shafqat Ul},
   eprint={2011.08023},
   archivePrefix={arXiv},
   primaryClass={gr-qc},
   year={2021},
   month=Mar, pages={056} }

@article{rotatingbh2,
   title={Strong field gravitational lensing by hairy Kerr black holes},
   volume={103},
   ISSN={2470-0029},
   url={http://dx.doi.org/10.1103/PhysRevD.103.124052},
   DOI={10.1103/physrevd.103.124052},
   number={12},
   journal={Physical Review D},
   publisher={American Physical Society (APS)},
   author={Islam, Shafqat Ul and Ghosh, Sushant G.},
   year={2021},
   month={June} }

@article{regularbh1,
   title={Strong gravitational lensing by rotating Simpson-Visser black holes},
   volume={2021},
   ISSN={1475-7516},
   url={http://dx.doi.org/10.1088/1475-7516/2021/10/013},
   DOI={10.1088/1475-7516/2021/10/013},
   number={10},
   journal={Journal of Cosmology and Astroparticle Physics},
   publisher={IOP Publishing},
   author={Islam, Shafqat Ul and Kumar, Jitendra and Ghosh, Sushant G.},
   eprint={2104.00696},
   archivePrefix={arXiv},
   primaryClass={gr-qc},
   year={2021},
   month=Oct, pages={013} }

@article{regularbh2,
   title={Testing Strong Gravitational Lensing Effects of Supermassive Compact Objects with Regular Spacetimes},
   volume={938},
   ISSN={1538-4357},
   url={http://dx.doi.org/10.3847/1538-4357/ac912c},
   DOI={10.3847/1538-4357/ac912c},
   number={2},
   journal={The Astrophysical Journal},
   publisher={American Astronomical Society},
   author={Kumar, Jitendra and Ul Islam, Shafqat and Ghosh, Sushant G.},
   year={2022},
   month=Oct, pages={104},
   }

@misc{loop1,
      title={Strong Gravitational Lensing by Loop Quantum Gravity Motivated Rotating Black Holes and EHT Observations}, 
      author={Jitendra Kumar and Shafqat Ul Islam and Sushant G. Ghosh},
      year={2023},
      journal = { The European Physical Journal C},
      eprint={2305.04336},
      archivePrefix={arXiv},
      primaryClass={gr-qc},
      url={https://arxiv.org/abs/2305.04336}, 
      doi={https://doi.org/10.1140/epjc/s10052-023-12205-3},
}

@misc{rotatingbh4,
      title={Probing Lorentz Symmetry Violation through Lensing Observables of Rotating Black Holes}, 
      author={Arun Kumar and Shafqat Ul Islam and Sushant G. Ghosh},
      year={2025},
      eprint={2509.00127},
      archivePrefix={arXiv},
      primaryClass={gr-qc},
      url={https://arxiv.org/abs/2509.00127}, 
}

@article{10.1098/rspa.1959.0015,
    author = {Darwin, Charles Galton},
    title = {The gravity field of a particle},
    journal = {Proceedings of the Royal Society of London. A. Mathematical and Physical Sciences},
    volume = {249},
    number = {1257},
    pages = {180-194},
    year = {1959},
    month = {01},
    issn = {0080-4630},
    doi = {10.1098/rspa.1959.0015},
    url = {https://doi.org/10.1098/rspa.1959.0015},
    eprint = {https://royalsocietypublishing.org/rspa/article-pdf/249/1257/180/51425/rspa.1959.0015.pdf},
}

@ARTICLE{1979A&A....75..228L,
       author = {{Luminet}, J.-P.},
        title = "{Image of a spherical black hole with thin accretion disk.}",
      journal = {\aap},
         year = 1979,
        month = may,
       volume = {75},
        pages = {228-235},
       url = {https://ui.adsabs.harvard.edu/abs/1979A&A....75..228L}
}

@article{PhysRevD.61.064021,
  title = {Spacetime perspective of Schwarzschild lensing},
  author = {Frittelli, Simonetta and Kling, Thomas P. and Newman, Ezra T.},
  journal = {Phys. Rev. D},
  volume = {61},
  issue = {6},
  pages = {064021},
  numpages = {14},
  year = {2000},
  month = {Feb},
  publisher = {American Physical Society},
  doi = {10.1103/PhysRevD.61.064021},
  url = {https://link.aps.org/doi/10.1103/PhysRevD.61.064021}
}

@article{Werner:2012rc,
    author = "Werner, M. C.",
    title = "{Gravitational lensing in the Kerr-Randers optical geometry}",
    journal = "Gen. Rel. Grav.",
    volume = "44",
    pages = "3047--3057",
    year = "2012",
    doi = "10.1007/s10714-012-1458-9"
}

@article{AyonBeato:1998ub,
    author = "Ayon-Beato, Eloy and Garcia, Alberto",
    title = "{Regular black hole in general relativity coupled to nonlinear electrodynamics}",
    journal = "Phys. Rev. Lett.",
    volume = "80",
    pages = "5056--5059",
    year = "1998",
    eprint = "gr-qc/9911046"
}

@article{Penrose:1969pc,
    author = "Penrose, R.",
    title = "{Gravitational collapse: The role of general relativity}",
    journal = "Riv. Nuovo Cim.",
    volume = "1",
    pages = "252--276",
    year = "1969"
}

@article{Carballo-Rubio:2018pmi,
    author = "Carballo-Rubio, Ra\'ul and Di Filippo, Francesco and Liberati, Stefano and Pacilio, Costantino and Visser, Matt",
    title = "{Inner horizon instability and the unstable core of regular black holes}",
    journal = "JHEP",
    volume = "05",
    pages = "132",
    year = "2018",
    eprint = "1802.10226"
}

@article{Brown:2011tv,
    author = "Brown, E. G. and Mann, R. B. and Modesto, L.",
    title = "{Mass Inflation in the Loop Black Hole}",
    journal = "Phys. Rev. D",
    volume = "84",
    pages = "104041",
    year = "2011",
    eprint = "1104.3126"
}

@article{Hsiao:2026oti,
    author = "Hsiao, Chen-Hung and Yuan, Limei and Wan, Yidun",
    title = "{Gravitational Lensing Signatures of Hayward-like Black Holes}",
    journal = "arXiv e-prints",
    pages = "arXiv:2604.14505",
    eprint = "2604.14505",
    archivePrefix = "arXiv",
    primaryClass = "gr-qc",
    month = "4",
    year = "2026"
}

@article{Yuan:2026bardeen,
    author = "Yuan, Limei and Hsiao, Chen-Hung and Wan, Yidun",
    title = "{Distinguish Bardeen-like black holes by Gravitational lensing}",
    journal = "J. High Energy Astrophys.",
    volume = "55",
    pages = "100713",
    year = "2027",
    eprint = "2604.11951",
    archivePrefix = "arXiv",
    primaryClass = "gr-qc",
    doi = "10.1016/j.jheap.2026.100713"
}

@article{Newman:1965tx,
    author = "Newman, E. T. and Janis, A. I.",
    title = "{Note on the Kerr spinning-particle metric}",
    journal = "J. Math. Phys.",
    volume = "6",
    pages = "915--917",
    year = "1965",
    doi = "10.1063/1.1704350"
}

@article{Azreg-Ainou:2014aqa,
    author = {Azreg-A{\"\i}nou, Mustapha},
    title = "{From static to rotating to conformal static solutions: Rotating imperfect fluid wormholes with(out) electric or magnetic field}",
    eprint = "1401.4292",
    archivePrefix = "arXiv",
    primaryClass = "gr-qc",
    doi = "10.1140/epjc/s10052-014-2865-8",
    journal = "Eur. Phys. J. C",
    volume = "74",
    number = "5",
    pages = "2865",
    year = "2014"
}

@article{Azreg-Ainou:2014pra,
    author = {Azreg-A{\"\i}nou, Mustapha},
    title = "{Generating rotating regular black hole solutions without complexification}",
    eprint = "1405.2569",
    archivePrefix = "arXiv",
    primaryClass = "gr-qc",
    doi = "10.1103/PhysRevD.90.064041",
    journal = "Phys. Rev. D",
    volume = "90",
    number = "6",
    pages = "064041",
    year = "2014"
}

@book{Hawking:1973uf,
    author = "Hawking, Stephen W. and Ellis, George F. R.",
    title = "{The Large Scale Structure of Space-Time}",
    doi = "10.1017/9781009253161",
    isbn = "978-1-009-25316-1, 978-1-009-25315-4, 978-0-521-20016-5",
    publisher = "Cambridge University Press",
    series = "Cambridge Monographs on Mathematical Physics",
    month = "2",
    year = "2023"
}

@article{Kontou:2020bta,
    author = "Kontou, Eleni-Alexandra and Sanders, Ko",
    title = "{Energy Conditions in General Relativity and Quantum Field Theory}",
    eprint = "2003.01815",
    archivePrefix = "arXiv",
    primaryClass = "gr-qc",
    doi = "10.1088/1361-6382/ab8fcf",
    journal = "Class. Quant. Grav.",
    volume = "37",
    number = "19",
    pages = "193001",
    year = "2020"
}

@article{Bambi:2013ufa,
    author = "Bambi, Cosimo and Modesto, Leonardo",
    title = "{Rotating regular black holes}",
    eprint = "1302.6075",
    archivePrefix = "arXiv",
    primaryClass = "gr-qc",
    doi = "10.1016/j.physletb.2013.03.025",
    journal = "Phys. Lett. B",
    volume = "721",
    pages = "329--334",
    year = "2013"
}

@article{Neves:2014aba,
    author = "Neves, J. C. S. and Saa, Alberto",
    title = "{Regular rotating black holes and the weak energy condition}",
    eprint = "1402.2694",
    archivePrefix = "arXiv",
    primaryClass = "gr-qc",
    doi = "10.1016/j.physletb.2014.05.026",
    journal = "Phys. Lett. B",
    volume = "734",
    pages = "44--48",
    year = "2014"
}

@article{Drake:1998gf,
    author = "Drake, S. P. and Szekeres, Peter",
    title = "{An explanation of the Newman-Janis algorithm}",
    eprint = "gr-qc/9807001",
    archivePrefix = "arXiv",
    journal = "Gen. Rel. Grav.",
    volume = "32",
    pages = "445--458",
    year = "2000"
}

@article{Gurses:1975vu,
    author = {G\"{u}rses, Metin and G\"{u}rsey, Feza},
    title = "{Lorentz covariant treatment of the Kerr-Schild geometry}",
    doi = "10.1063/1.522480",
    journal = "J. Math. Phys.",
    volume = "16",
    pages = "2385--2390",
    year = "1975"
}

@article{Toshmatov:2014nya,
    author = {Toshmatov, Bobir and Ahmedov, Bobomurat and Abdujabbarov, Ahmadjon and Stuchl\'ik, Zden\v{e}k},
    title = "{Rotating regular black hole solution}",
    eprint = "1404.6443",
    archivePrefix = "arXiv",
    primaryClass = "gr-qc",
    doi = "10.1103/PhysRevD.89.104017",
    journal = "Phys. Rev. D",
    volume = "89",
    number = "10",
    pages = "104017",
    year = "2014"
}

@article{Torres:2016gpe,
    author = "Torres, Ram\'on and Fayos, Francesc",
    title = "{On regular rotating black holes}",
    eprint = "1611.03654",
    archivePrefix = "arXiv",
    primaryClass = "gr-qc",
    journal = "Gen. Rel. Grav.",
    volume = "49",
    number = "1",
    pages = "2",
    year = "2017"
}

@incollection{Torres:2022regularreview,
    author = "Torres, Ram{\'o}n",
    title = "{Regular Rotating Black Holes}",
    booktitle = "{Regular Black Holes: Towards a New Paradigm of Gravitational Collapse}",
    editor = "Bambi, Cosimo",
    publisher = "Springer Singapore",
    pages = "421--446",
    year = "2023",
    eprint = "2208.12713",
    archivePrefix = "arXiv",
    primaryClass = "gr-qc",
    doi = "10.1007/978-981-99-1596-5_11"
}

@article{Carter:1968separable,
    author = "Carter, Brandon",
    title = "{Hamilton-Jacobi and Schrodinger Separable Solutions of Einstein's Equations}",
    doi = "10.1007/BF03399503",
    journal = "Commun. Math. Phys.",
    volume = "10",
    pages = "280--310",
    year = "1968"
}

@article{Benenti:1979separability,
    author = "Benenti, Sergio and Francaviglia, Mauro",
    title = "{Remarks on Certain Separability Structures and Their Applications to General Relativity}",
    doi = "10.1007/BF00757025",
    journal = "Gen. Rel. Grav.",
    volume = "10",
    pages = "79--92",
    year = "1979"
}

@book{Chandrasekhar:1983blackholes,
    author = "Chandrasekhar, Subrahmanyan",
    title = "{The Mathematical Theory of Black Holes}",
    publisher = "Clarendon Press",
    address = "Oxford",
    year = "1983"
}

@inproceedings{Bardeen:1973shadow,
    author = "Bardeen, James M.",
    title = "{Timelike and Null Geodesics in the Kerr Metric}",
    editor = "DeWitt, C. and DeWitt, B. S.",
    booktitle = "{Black Holes (Les Astres Occlus)}",
    publisher = "Gordon and Breach",
    address = "New York",
    pages = "215--240",
    year = "1973"
}

@article{Teo:2003bfn,
    author = "Teo, Edward",
    title = "{Spherical Photon Orbits Around a Kerr Black Hole}",
    doi = "10.1023/A:1026286607562",
    journal = "Gen. Rel. Grav.",
    volume = "35",
    number = "11",
    pages = "1909--1926",
    year = "2003"
}

@article{Grenzebach:2014fha,
    author = "Grenzebach, Arne and Perlick, Volker and L{\"a}mmerzahl, Claus",
    title = "{Photon Regions and Shadows of Kerr-Newman-NUT Black Holes with a Cosmological Constant}",
    eprint = "1403.5234",
    archivePrefix = "arXiv",
    primaryClass = "gr-qc",
    doi = "10.1103/PhysRevD.89.124004",
    journal = "Phys. Rev. D",
    volume = "89",
    number = "12",
    pages = "124004",
    year = "2014"
}

@article{Perlick:2021aok,
    author = "Perlick, Volker and Tsupko, Oleg Yu.",
    title = "{Calculating Black Hole Shadows: Review of Analytical Studies}",
    eprint = "2105.07101",
    archivePrefix = "arXiv",
    primaryClass = "gr-qc",
    doi = "10.1016/j.physrep.2021.10.004",
    journal = "Phys. Rept.",
    volume = "947",
    pages = "1--39",
    year = "2022"
}

@article{Komar:1959,
    author = "Komar, Arthur",
    title = "{Covariant Conservation Laws in General Relativity}",
    doi = "10.1103/PhysRev.113.934",
    journal = "Phys. Rev.",
    volume = "113",
    pages = "934--936",
    year = "1959"
}

@book{Wald:1984gr,
    author = "Wald, Robert M.",
    title = "{General Relativity}",
    publisher = "University of Chicago Press",
    address = "Chicago",
    year = "1984"
}

@article{Ali:2024rqcbh,
    author = "Ali, Heena and Islam, Shafqat Ul and Ghosh, Sushant G.",
    title = "{Shadows and Parameter Estimation of Rotating Quantum-Corrected Black Holes and Constraints from EHT Observation of M87* and Sgr A*}",
    eprint = "2410.09198",
    archivePrefix = "arXiv",
    primaryClass = "gr-qc",
    doi = "10.1016/j.jheap.2025.100367",
    journal = "JHEAp",
    volume = "47",
    pages = "100367",
    year = "2025"
}

@article{Hioki:2009shadow,
    author = "Hioki, Kenta and Maeda, Kei-ichi",
    title = "{Measurement of the Kerr Spin Parameter by Observation of a Compact Object's Shadow}",
    eprint = "0904.3575",
    archivePrefix = "arXiv",
    primaryClass = "astro-ph.HE",
    doi = "10.1103/PhysRevD.80.024042",
    journal = "Phys. Rev. D",
    volume = "80",
    pages = "024042",
    year = "2009"
}

@article{Bambi:2019M87,
    author = "Bambi, Cosimo and Freese, Katherine and Vagnozzi, Sunny and Visinelli, Luca",
    title = "{Testing the Rotational Nature of the Supermassive Object M87* from the Circularity and Size of Its First Image}",
    eprint = "1904.12983",
    archivePrefix = "arXiv",
    primaryClass = "gr-qc",
    doi = "10.1103/PhysRevD.100.044057",
    journal = "Phys. Rev. D",
    volume = "100",
    number = "4",
    pages = "044057",
    year = "2019"
}

@article{Afrin:2021hairyKerr,
    author = "Afrin, Misba and Kumar, Rahul and Ghosh, Sushant G.",
    title = "{Parameter Estimation of Hairy Kerr Black Holes from Its Shadow and Constraints from M87*}",
    eprint = "2103.11417",
    archivePrefix = "arXiv",
    primaryClass = "gr-qc",
    doi = "10.1093/mnras/stab1260",
    journal = "Mon. Not. Roy. Astron. Soc.",
    volume = "504",
    pages = "5927--5940",
    year = "2021"
}

@article{Xie:2024srr,
    author = "Xie, Chen-Hao and Zhang, Yu and Sun, Qi and Li, Qi-Quan and Duan, Peng-Fei",
    title = "{Gravitational Lensing by a Stable Rotating Regular Black Hole}",
    eprint = "2401.05454",
    archivePrefix = "arXiv",
    primaryClass = "gr-qc",
    doi = "10.1088/1475-7516/2024/05/121",
    journal = "JCAP",
    volume = "05",
    pages = "121",
    year = "2024"
}

@article{Guo:2025rrbl,
    author = "Guo, Ming-Yu and Wu, Meng-He and Guo, Hong and Kuang, Xiao-Mei and Liu, Fu-Yao",
    title = "{Strong Gravitational Lensing Effects around Rotating Regular Black Holes}",
    eprint = "2501.00292",
    archivePrefix = "arXiv",
    primaryClass = "gr-qc",
    doi = "10.1016/j.physletb.2024.139211",
    journal = "Phys. Lett. B",
    volume = "860",
    pages = "139211",
    year = "2025"
}

@article{Nengroo:2026kalbRamond,
    author = "Nengroo, Towheed Ahmad and Islam, Shafqat Ul and Ghosh, Sushant G.",
    title = "{Probing Kalb--Ramond Gravity with Charged Rotating Black Holes: Constraints from EHT Observations}",
    eprint = "2604.13494",
    archivePrefix = "arXiv",
    primaryClass = "gr-qc",
    doi = "10.1016/j.dark.2026.102339",
    journal = "Phys. Dark Univ.",
    volume = "52",
    pages = "102339",
    year = "2026"
}

@article{Kumar:2020shadowParameters,
    author = "Kumar, Rahul and Ghosh, Sushant G.",
    title = "{Black Hole Parameter Estimation from Its Shadow}",
    eprint = "1811.01260",
    archivePrefix = "arXiv",
    primaryClass = "gr-qc",
    doi = "10.3847/1538-4357/ab77b0",
    journal = "Astrophys. J.",
    volume = "892",
    number = "2",
    pages = "78",
    year = "2020"
}

@article{Afrin:2024khy,
    author = "Afrin, Misba and Ghosh, Sushant G. and Wang, Anzhong",
    title = "{Testing EGB gravity coupled to bumblebee field and black hole parameter estimation with EHT observations}",
    eprint = "2409.06218",
    archivePrefix = "arXiv",
    primaryClass = "gr-qc",
    doi = "10.1016/j.dark.2024.101642",
    journal = "Phys. Dark Univ.",
    volume = "46",
    pages = "101642",
    year = "2024"
}
\bibliographystyle{aasjournalv7}

\end{document}